\documentclass[journal, letterpaper]{IEEEtran}
\usepackage[cmex10]{amsmath}
\usepackage{amssymb}
\usepackage{graphicx}
\usepackage[caption=false,font=footnotesize,farskip=0.2em]{subfig}
\usepackage{listings}
\usepackage{url}
\usepackage{multirow}
\usepackage{hhline}
\usepackage{nicefrac}
\usepackage{lipsum}
\usepackage{balance}
\usepackage{cite}
\IEEEoverridecommandlockouts
\begin{document}
\title{A Benchmark for the Performance of Time-varying Closed-loop Flow Control with Application to TCP}
\author{\IEEEauthorblockN{Ralf L\"{u}bben and Markus Fidler}
\thanks{R. L\"{u}bben and M. Fidler are with the Institute of Communications Technology, Leibniz Universit\"{a}t Hannover.}
\thanks{This work was supported in part by the German Research Foundation (DFG) under Grant NetMeter (FI 1236/2-2).}
\thanks{This manuscript is an extended version of the paper~\cite{luebben:TCPscest} that appeared in the IEEE Infocom 2016 proceedings.}
}
\maketitle
\begin{abstract}
Closed-loop flow control protocols, such as the prominent implementation TCP, are prevalent in the Internet, today. TCP has continuously been improved for greedy traffic sources to achieve high throughput over networks with large bandwidth delay products. Recently, the increasing use for streaming and interactive applications, such as voice and video, has shifted the focus towards its delay performance. Given the need for real-time communication of non-greedy sources via TCP, we present an estimation method for performance evaluation of closed-loop flow control protocols. We characterize an end-to-end connection by a transfer function that provides statistical service guarantees for arbitrary traffic. The estimation is based on end-to-end measurements at the application level that include all effects induced by the network and by the protocol stacks of the end systems. From our measurements, we identify different causes for delays. We show that significant delays are due to queueing in protocol stacks. Notably, this occurs even if the utilization is moderate. Using our estimation method, we compare the impact of fundamental mechanisms of TCP on delays at the application level: In detail, we analyze parameters relevant for network dimensioning, including buffer provisioning and active queue management, and parameters for server configuration, such as the congestion control algorithm. By applying our method as a benchmark, we find that a good selection can largely improve the delay performance of TCP.
\end{abstract}
\section{Introduction}
\label{sec:intro}
In today's Internet, closed-loop flow control protocols are prevalent. The most prominent implementation is the transmission control protocol (TCP). Besides preventing congestion, throughput and fairness of TCP data flows are an active area of research since many years. Particularly, the increase of bandwidth and hence of the bandwidth delay product (BDP) has motivated significant works that led to new TCP congestion control algorithms. Examples are Compound~\cite{Compound} and Cubic~\cite{Ha:Cubic} that prevail at present~\cite{Yang:2011}. These congestion control algorithms aim at achieving a high utilization of network paths with a large BDP, assuming that sources are greedy.

Today, numerous applications exist that have a significantly different profile: delay-sensitive applications such as interactive voice and video or streaming video require small to moderate end-to-end delays in the range of 100~ms up to a few seconds, respectively. These application have rather rigid bandwidth requirements from about one hundred kbit/s up to several Mbit/s. They account for more than 60\% of the data traffic in fixed access networks and more than 30\% in mobile access networks as reported in~\cite{sandvine}. Many of these applications rely on HTTP and thus on TCP for data transport as identified in~\cite{Maier:InternetTraffic} to benefit from the existing infrastructure of content delivery networks, which also reduce the round trip time significantly~\cite{akamaiRTT2009}. Aside of HTTP, TCP is often used as a fall-back transport protocol by real-time applications, such as video telephony, if UDP is blocked by firewalls~\cite{Xu:videotel,rfc5766}. Even if UDP is used, closed-loop flow control may be integrated on top, e.g., the protocol Quick UDP Internet Connections mimics TCP's congestion control~\cite{quic_ietf}.

Up to now, large client-side buffers are used for streaming of recorded content to mitigate effects of variable bandwidth. Well-known streaming services buffer several tens of seconds at the client side~\cite{Rao:streaming_characteristic}. With the same goal but without extensive buffering, adaptive streaming techniques are developed as, e.g., dynamic adaptive streaming over HTTP~\cite{DeCicco:dash} that adapt the data rate to the available bandwidth and client capabilities. This is also applicable for live-streaming since extensive buffering is not required.

This exemplary list illustrates the importance of considering the delay performance of flow-control protocols for delay sensitive and non-greedy applications. Yet, methods for performance evaluation focus mainly on metrics aligned to primal applications as fairness, long-term throughput, or if at all on the completion time of data flows. Regarding an evaluation of delay, which occurs while a connection is active, for any type of application traffic, no coherent method exists.

In this paper, we present a method to study the delay performance of closed-loop flow control protocols. The method provides an estimate of a transfer function of an end-to-end path from measurements and is based on a system model of closed-loop window flow control. The transfer function enables, among others, the computation of delay bounds that hold for the entire duration of a connection. We denote the transfer function as service process $S_{app}(\tau,t)$ for $t \ge \tau \ge 0$ that specifies the service available between times $\tau$ and $t$. The service process $S_{app}(\tau,t)$ maps the cumulative traffic arrivals $A_{app}(t)$ that are generated by the sender's application up to time $t$ to the departures $D_{app}(t)$ that are delivered to the receiver's application. The estimate of $S_{app}(\tau,t)$ is constructed from quantiles of the steady-state delay distribution that is observed for the connection by constant rate probe traffic. Compared to the related work that typically uses greedy sources, the function $S_{app}(\tau,t)$ enables the computation of statistical guarantees for traffic departures, backlog, and delay for arbitrary application traffic.

\begin{figure}[t]
\centering
  \includegraphics[width=0.8\columnwidth]{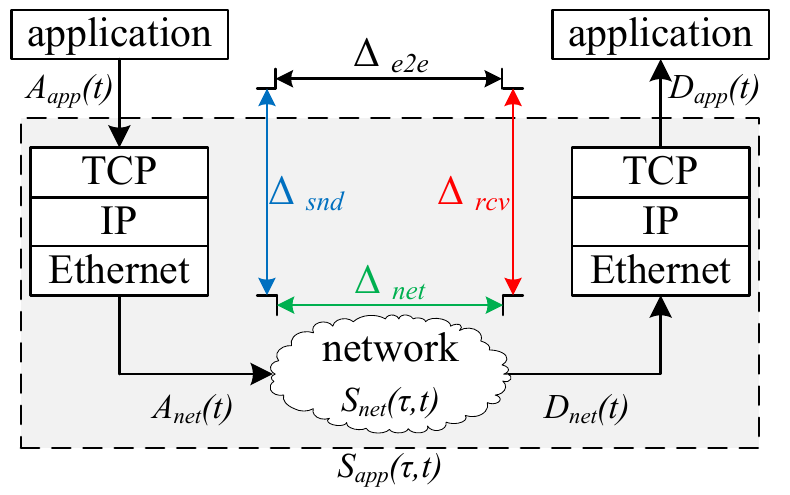}
   \caption{The path that application data traverse from the sender to the receiver is composed of three segments: the sender's protocol stack, the network, and the receiver's protocol stack. To the application, the TCP connection appears as a transparent Byte stream that is modeled by a service process $S_{app}(\tau,t)$.}
  \label{fig:abstraction_network_path}
\end{figure}

Fig.~\ref{fig:abstraction_network_path} illustrates this end-to-end view for the case of TCP: the service process $S_{app}(\tau,t)$ considers the service that a connection offers to the application, i.e., the service that is available between the socket at the sending application and the socket at the receiving application. This path is composed of three segments: the sender's protocol stack, the network, and the receiver's protocol stack. Noteworthy, this perspective provides a holistic view on the complete path. It includes effects of the network path as well as the transport protocols at the sender and receiver side. This enables us to determine the impact of the configuration and the properties of such segments, as the influence of queue management or the congestion control algorithm. For wireless networks, such a view is effectively used for performance evaluation~\cite{wu:effcap}. A similar perspective is considered by the flow completion time as promoted in~\cite{dukkipati2006flow}. Beyond the flow completion time, the notion of service process reveals delays that occur while a flow is active. Such delays are negligible for downloads but crucial for real-time and streaming applications.

To exemplify the relevance of having a description such as $S_{app}(\tau,t)$ that includes the network as well as the protocol stacks of the end systems, we present first delay measurements. The delays are a result of the variability of the service process $S_{app}(\tau,t)$. We show a decomposition of the end-to-end delay $\Delta_{e2e}$ of a tagged TCP connection into the delay components $\Delta_{snd}$, $\Delta_{net}$, and $\Delta_{rcv}$ that occur in the sender's protocol stack, the network, and the receiver's stack, respectively, see Fig.~\ref{fig:abstraction_network_path}. We consider a network with a 100~Mbit/s bottleneck link that is loaded by nine other greedy TCP cross traffic flows (not depicted in Fig.~\ref{fig:abstraction_network_path}). As a consequence, the network capacity is generally highly saturated with TCP traffic, where the tagged TCP connection is eligible for a fair share of 10\% of the network capacity, i.e., about 10~Mbit/s. More details on the setup are provided in Sec.~\ref{sec:exp_setup}.

\begin{figure}
 \centering
  \subfloat[ low utilization (1~Mbit/s) ]{
    \includegraphics[width=0.48\columnwidth]{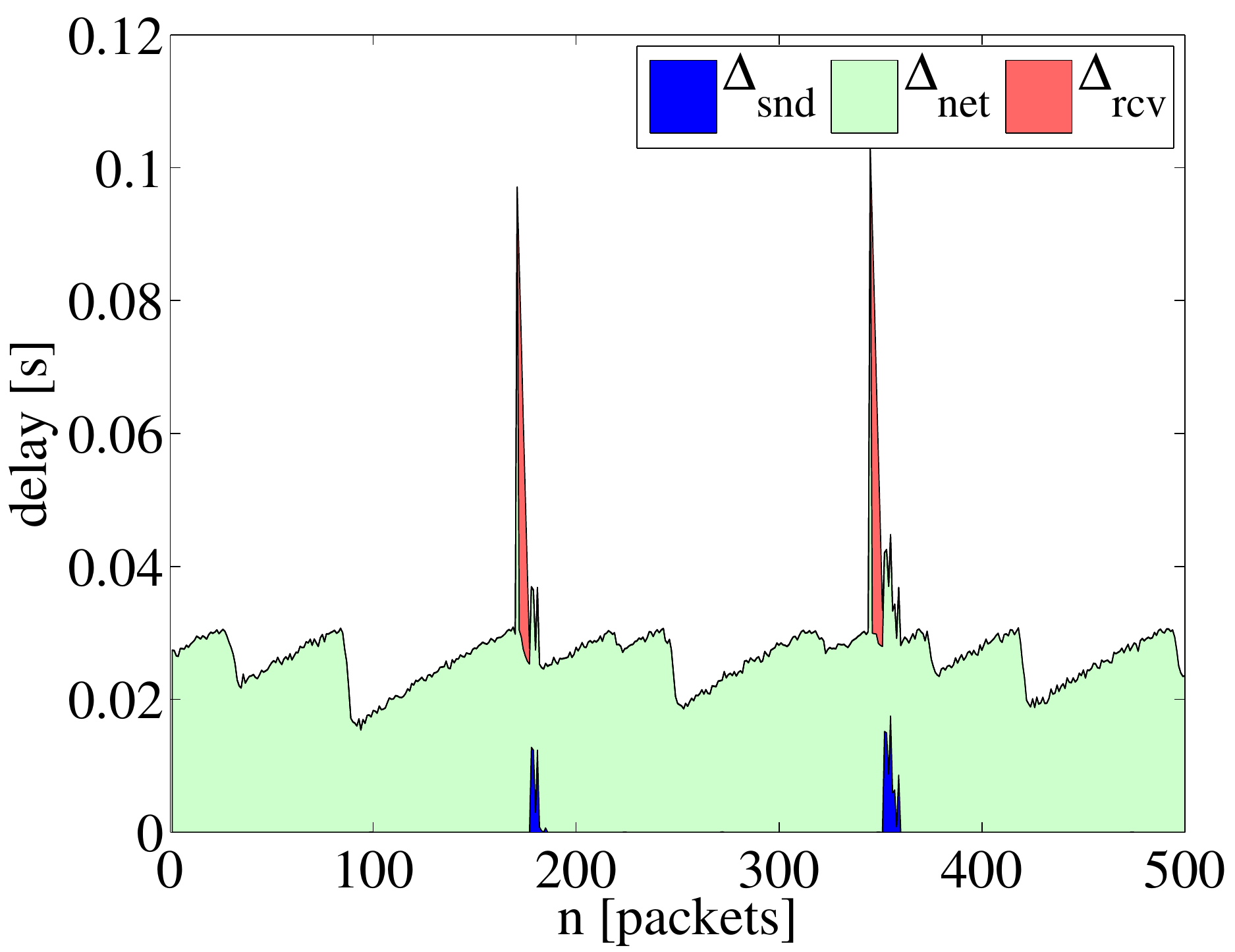}
    \label{fig:tcp_queue_size_delay_1}
  }
  \subfloat [ high utilization (9~Mbit/s) ]{
    \includegraphics[width=0.48\columnwidth]{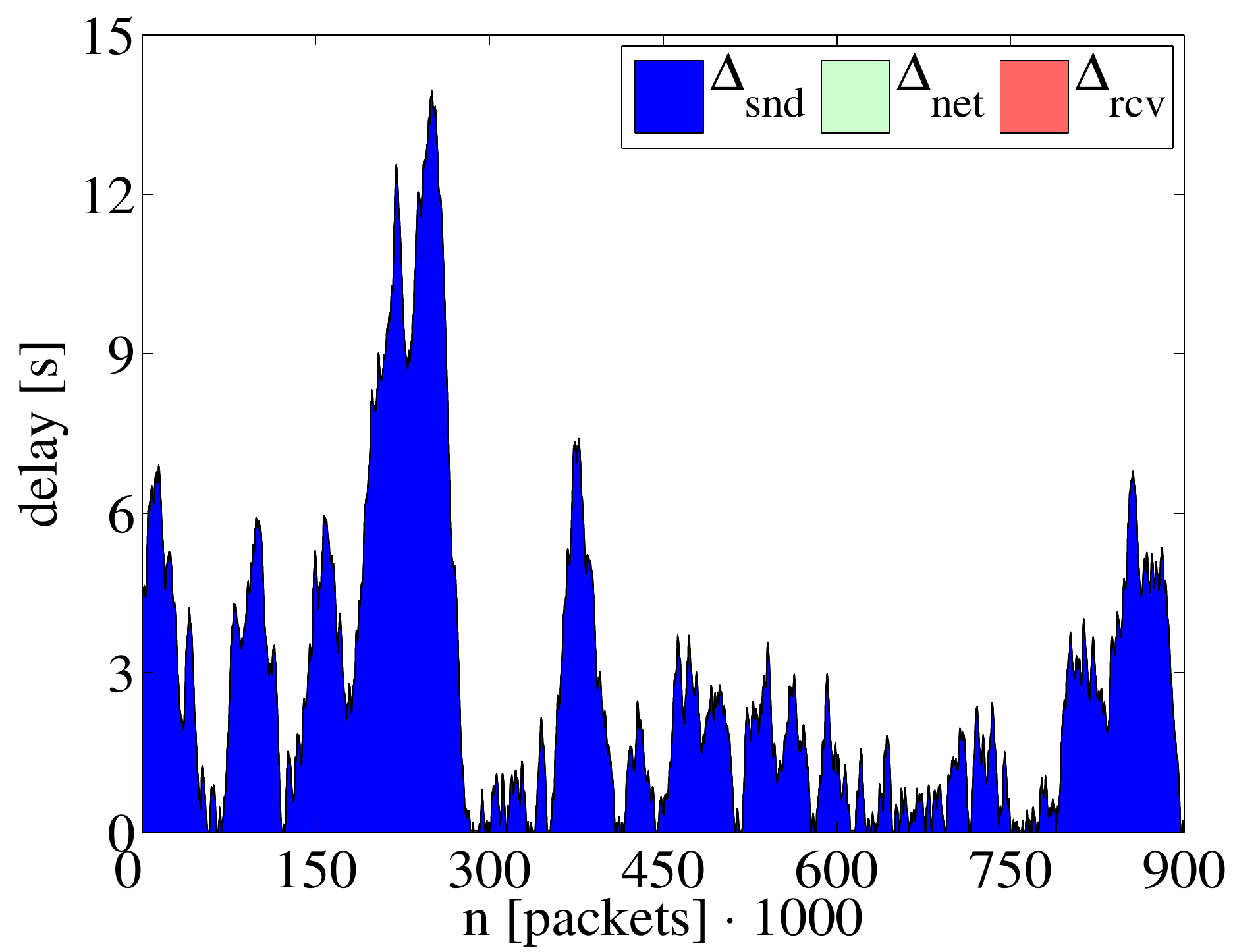}
    \label{fig:tcp_queue_size_delay_9}
  }
  \caption{Delay components $\Delta_{snd}$, $\Delta_{net}$, and $\Delta_{rcv}$. In case of low utilization, Fig.~\ref{fig:tcp_queue_size_delay_1}, delays due to propagation and queueing in the network are prevalent. Occasionally, further delays occur due to retransmissions and to recover in-order delivery of packets. In case of high utilization, Fig.~\ref{fig:tcp_queue_size_delay_9}, large delays in the sender's protocol stack dominate and last for long periods.}
  \label{fig:cp_queue_size_delay}
\end{figure}

In the stacked graph in Fig.~\ref{fig:tcp_queue_size_delay_1}, we present the individual parts of the delay $\Delta_{e2e}$ of the tagged TCP connection that are observed for a constant application data rate $A_{app}(t) = rt$ of $r=1$~Mbit/s. This rate corresponds to a low utilization of the TCP connection with respect to its fair capacity share. Note that the network nevertheless is highly loaded due to the greedy TCP cross-traffic flows. In this experiment, network delays, that are composed of a constant propagation delay of 10~ms and a variable queueing delay of at most 20~ms, are prevalent. Two peaks in the delay series of about 100~ms occur after a packet loss. They comprise the initial transmission ($\le$~30~ms), a latency until three duplicate acknowledgements are received ($\approx$ 46 ms), and the retransmission ($\le$ 30 ms). Note that subsequent packets, that are successfully received after the loss, also incur additional delays as they have to wait at the receiver for the retransmission to ensure in-order delivery. Delays due to queueing in the sender's protocol stack occur rarely and are mostly small.

In case of an application data rate of 9~Mbit/s, that corresponds to a high utilization of the tagged TCP connection, end-to-end delays increase drastically, as shown in Fig.~\ref{fig:tcp_queue_size_delay_9}. Delays in the range of several seconds occur in the sender's protocol stack already in this single bottleneck network topology. They originate from queueing inside the protocol stack or blocking if the socket buffer is full. This is due to TCP if it emits packets with a rate less than the application rate, which is a consequence of TCP's congestion control. These delays can last for a large number of subsequent packets. The other delay components are hardly visible since they are dominated by the delays in the sender's stack.

Our first results highlight the significance of delays that occur in protocol stacks, e.g., as a consequence of packet loss and congestion control. These stack delays can exceed network delays by orders of magnitude. TCP's control loop results in a complex interaction with the network. How modifications of, e.g., the TCP congestion control algorithm or the queue management, influence the overall performance is a hard problem. To approach this problem, we propose a method for estimation of the service process $S_{app}(\tau,t)$ from measurements based on a system model for closed-loop flow control connections. Using the method we provide relevant insights into the question of the performance of TCP and the impact of fundamental parameters.

We apply our method to evaluate the end-to-end performance of TCP and contribute an analysis of the end-to-end delay distribution of TCP connections.
Firstly, we discover distinct effects that are due to queueing in the network and in the sender's protocol stack, respectively. As a result of limited buffers at routers, the distribution of delays that are caused by queueing in the network decays faster than exponentially. In contrast, delays that are caused by queueing in the protocol stack exhibit a slower, approximately exponential decay, and account largely for the tail distribution. We analyze the reasons for delays and show how the temporal characteristics of congestion control take a decisive effect.
Secondly, we use estimates of the service processes of TCP connections for a comparative evaluation, where we consider the impact of basic parameters: on the one hand, network configurations as the buffer size at routers or the active queue management (AQM), on the other hand, parameters configurable at application servers as the application rate and the congestion control algorithm. Our results reveal that certain configurations can reduce end-to-end delays substantially by about 50\%.

The remainder of this paper is structured as follows. In Sec.~\ref{sec:related_work}, we present the related work. Sec.~\ref{sec:pref_eval} provides the system model for closed-loop window flow control and the estimation method. The experimental setup is introduced in Sec.~\ref{sec:exp_setup}. In Sec.~\ref{sec:eval_reno}, we examine in detail the service that TCP NewReno provides to an application and explain its characteristics by decomposition of the end-to-end delays with respect to different root causes. Sec.~\ref{sec:benchmark2} presents a benchmark for the impact of basic parameters on TCP's service including utilization, buffer size, queue management, and congestion control algorithm. For each of the parameters, we discuss the impacts on TCP's performance. Sec.~\ref{sec:comparison} continues the benchmark demonstrating the performance impact from an application's perspective. Sec.~\ref{sec:conclusion} concludes the paper.
\section{Related Work}
\label{sec:related_work}
A large body of literature exists on the evaluation of the throughput of closed-loop flow control protocols and its prevalent realization TCP. Here, we focus on the works that go beyond throughput evaluation and provide results in terms of the delay performance or service processes as introduced in Sec.~\ref{sec:intro}. Regarding latency reduction techniques, that are not in the focus of our work, we refer the reader to the comprehensive survey~\cite{rite}.

Models of closed-loop flow control by a transfer function, as introduced in Sec.~\ref{sec:intro}, are derived in~\cite{Agrawal:flow_control, chang:performanceguarantees, leboudec:networkcalculus, liebeherr:windowflowcontrol, schmitt:windowflowcontrol}. The model in~\cite{Agrawal:flow_control, chang:performanceguarantees, leboudec:networkcalculus} enables the derivation of guarantees for throughput as well as delay for arbitrary arrival traffic. A shortcoming of the approach is, however, the assumption of a deterministic system. Random effects, such as a time-varying network service or an adaptive window size, are expressed by deterministic bounding functions using limiting values, e.g., of the window size. This reduces complexity and allows for an analytical solution. Window flow control systems with a time-varying service have been identified as a hard problem~\cite{Book-Jiang,survey_netcalc} and only recently first solutions~\cite{liebeherr:windowflowcontrol, schmitt:windowflowcontrol} have been derived. How these works can be extended to flow control with an adaptive window as in case of TCP is still to be investigated.

As TCP is the prevalent closed-loop flow control protocol, the related work on performance evaluation is manifold, especially with respect to the throughput. Here, we consider the less extensively studied area of the delay performance. The flow-completion time is modeled in~\cite{Cardwell:TCPLatency} considering TCP's initial startup effects. TCP for streaming applications is analyzed in~\cite{Wang:tcp_streaming}, the results indicate that the TCP flow should be eligible for twice the rate of the streaming application for proper streaming. Furthermore, the minimal receiver buffer size is derived analytically with respect to video quality, video traffic characteristics, and network characteristics in~\cite{kim2006recvbuftcp}. The last three models rely on the basic model of~\cite{Padhye:tcp}. In this model independent losses between rounds (back-to-back packets until first ACK) and TCP Reno are assumed. With these assumptions an accurate computation of the expected long-term throughput is feasible. The impact of more complex loss schemes and other congestion control protocols is, however, hardly mathematically tractable. For various TCP versions, steady-state distributions of queueing delays are considered in~\cite{wierman:markov_model} for on-off traffic sources. Based thereon, an analytical evaluation of the steady-state delay distribution of TCP NewReno, that includes the buffer size dynamics of the protocol stack, is provided for constant bit rate sources in~\cite{brosh:delayfriendliness}. Again, the model is tailored to specific TCP variants and independent losses.

To evaluate also scenarios in which dynamics are difficult to deal with, we bridge the two domains of measurements and analytical performance evaluation of networks with flow control. We develop a stochastic service curve model of closed-loop window flow control and use a method that estimates the shape of the service curve from measurements of real systems. Consequently, the results include all effects that are inherent in the network path and the protocol stacks, as buffer dynamics, which can be responsible for large delays due to congestion control as we show in Sec.~\ref{sec:eval_reno}. Due to the service curve property, the estimate can be used to provide performance guarantees for arbitrary traffic arrivals.
\section{Estimation of Closed-loop Flow Control}
\label{sec:pref_eval}
As highlighted in~\cite{Gettys:bufferbloat}, there is a need for new performance tests that evaluate metrics beyond TCP throughput. In this section, we present an estimation method that is able to characterize the end-to-end attainable rate of a TCP connection on arbitrary time-scales by a stochastic service curve. In the limit, this service curve converges to the long-term throughput. Moreover, it includes all aspects of end-to-end paths. We make use of this feature in Sec.~\ref{sec:benchmark2} to benchmark relevant parameters. The service curve is applicable for arbitrary traffic arrivals and enables the computation of statistical backlog and delay bounds $\mathsf{P} [ backlog > x ] \leq \varepsilon$ and $\mathsf{P} [ delay > x] \leq \varepsilon$, respectively, where $\varepsilon$ is a defined, small probability.

Firstly, we introduce a system model that enables a mathema\-ti\-cal formulation of the dynamics of closed-loop flow control. Secondly, we describe the estimation method and its implementation based on this model. The method makes use of constant rate packet train probes $A_{app} = rt$, i.e., packets are injected at a constant rate $r$ into a TCP socket to identify its end-to-end characteristics as depicted in Fig.~\ref{fig:abstraction_network_path}. Thirdly, we show estimates obtained for flow control with a static and an adaptive window, respectively, and compare these to analytical reference results. We show that the adaptation of the window size has a significant impact on the end-to-end service. We conclude this section with a numerical validation.
\subsection{System Model}
\label{sec:system_model}
We adopt a model of linear time-varying systems, referred to as exact dynamic server in~\cite{chang:performanceguarantees}. The model uses random service processes $S(\tau,t)$ to characterize the service offered by buffered systems, such as links and schedulers, in the time interval $(\tau,t]$. We denote the cumulative traffic arrivals and departures of a system in the interval $(0,t]$ by $A(t)$ and $D(t)$, respectively. By convention, there are no arrivals for $t \le 0$. We assume a fluid-flow model where data are infinitely divisible. Packetization effects can be included at the expense of additional notation. Generally, the deviation of the fluid system is at most one maximum packet length~\cite{chang:performanceguarantees, leboudec:networkcalculus}.
The arrival-departure relation of a system is given for $t \ge 0$ as~\cite{chang:performanceguarantees}
\begin{equation}
D(t)=\inf_{\tau \in [0,t]} \{A(\tau) + S(\tau,t) \}.
\label{equ:sc_time_varying}
\end{equation}

To give a first example, the service process of a constant rate link with capacity $C$ is $S(\tau,t)=C(t-\tau)$. If the link has in addition a constant delay $T$, the service process becomes a latency-rate function $S(\tau,t)=C[t-\tau-T]^+$, where $[x]^+\!=\!\max \{0,x \}$. Eq.~\eqref{equ:sc_time_varying} enables the computation of performance metrics, such as the backlog $B(t) = A(t) - D(t)$ for known but otherwise arbitrary arrivals. By iterative application, Eq.~\eqref{equ:sc_time_varying} extends immediately to tandem systems. For a broader introduction we refer to~\cite{chang:performanceguarantees, leboudec:networkcalculus, Book-Jiang, survey_netcalc}.

\begin{figure}
  \centering
  \includegraphics[width=0.7\columnwidth]{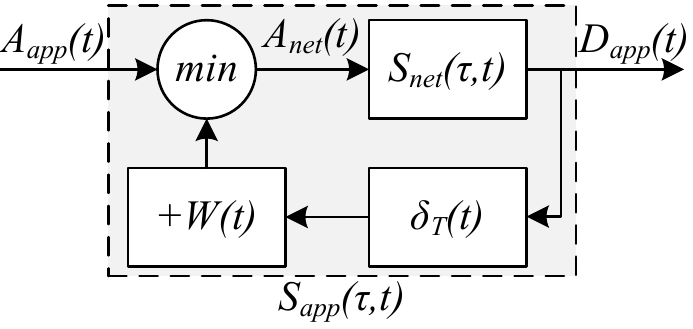}
  \caption{Arrivals to the network are regulated by closed-loop flow control.}
  \label{fig:window_flow_control_scheme}
\end{figure}
Next, we model a network path with closed-loop window flow control as illustrated in Fig.~\ref{fig:window_flow_control_scheme}. Due to, e.g., an adaptive window that changes over time, the end-to-end service of the flow control loop is random. Related models have been derived using linear dynamical systems~\cite{Baccelli:tcp_max_plus, chang:performanceguarantees}, for a static window in the framework of the deterministic network calculus~\cite{Agrawal:flow_control, chang:performanceguarantees, leboudec:networkcalculus}, and in the framework of the stochastic network calculus~\cite{liebeherr:windowflowcontrol, schmitt:windowflowcontrol}.

The forward path in Fig.~\ref{fig:window_flow_control_scheme} maps external application  data arrivals $A_{app}(t)$ to the departures $D_{app}(t)$. It consists of a throttle for flow control, i.e., the $min$ operator, that regulates the arrivals to the network $A_{net}(t)$, and the service process of the network $S_{net}(\tau,t)$. The reverse path is used by acknowledgements. For now, we assume it comprises a delay element $\delta_T(t)$ that causes a time shift of $T$. We model the service for acknowledgements as constant delay since acknowledgements are typically much smaller than data packets such that transmission times can be neglected. The time-varying window size is given by $W(t)$.

Tracking the structure in Fig.~\ref{fig:window_flow_control_scheme}, the departures for $t \ge 0$ are
\begin{equation}
D_{app}(t)= \inf_{\tau \in [0,t]} \{A_{net}(\tau) + S_{net}(\tau,t) \}
\label{eq:recursive_fc}
\end{equation}
from Eq.~\eqref{equ:sc_time_varying}, and the arrivals to the network for $t \ge 0$ are
\begin{equation}
A_{net}(t)= \min \left\{ A_{app}(t), D_{app}([t-T]^+) + W(t) \right\}.
\label{eq:recursive_fc_a}
\end{equation}
By recursive insertion of Eq.~\eqref{eq:recursive_fc_a} into Eq.~\eqref{eq:recursive_fc} a solution of the form $D_{app}(t) = \inf_{\tau \in [0,t]} \{A_{app}(\tau) + S_{app}(\tau,t)\}$ can be derived, i.e., window flow control has a service process $S_{app}(\tau,t)$ that satisfies Eq.~\eqref{equ:sc_time_varying}~\cite[p. 180]{chang:performanceguarantees}. Closed form solutions exist for certain cases. An example is flow control with a static window size $W(t) = w$ and a network with constant capacity $C$, latency $T$, and round trip time $RTT = 2T$, i.e., $S_{net}(\tau,t) = C [t-\tau-T]^+$. The fundamental insight in~\cite{Agrawal:flow_control} is that $S_{app}(\tau,t) = S_{net}(\tau,t)$ if $w \ge C \cdot RTT$, i.e., the network capacity can be fully utilized as long as the window is at least BDP-sized. Otherwise if $w < C \cdot RTT$, the long-term rate of $S_{app}(\tau,t)$ is limited by $w/RTT$. Recently, first solutions for window flow control with a random service are presented in~\cite{liebeherr:windowflowcontrol,schmitt:windowflowcontrol}. Both works emphasize on the difficulty to find a closed form solution for the recursion as already depicted in Eq.~\eqref{eq:recursive_fc} and Eq.~\eqref{eq:recursive_fc_a}. In the following, we present a method that enables us to retrieve a service estimate for the general class of closed loop window flow control from measurements. By use of measurements, we avoid the difficulty of an analytical solution, that may not be feasible given the complexity, e.g., of a full TCP implementation.
\subsection{Estimation Method}
\label{sec:probe_method}
We seek to obtain an estimate of the service process $S_{app}(\tau,t)$ of window flow control from measurements of $D_{app}(t)$ using selected probing traffic $A_{app}(t)$.
In the sequel, we characterize random service processes $S(\tau,t)$ from~\eqref{equ:sc_time_varying} by a statistical lower bound $\mathcal{S}^{\varepsilon}(t-\tau)$ that conforms with
\begin{equation}
\mathsf{P} \big[ S(\tau,t) \ge \mathcal{S}^{\varepsilon}(t-\tau) \;, \forall \tau \in [0,t] \big] \geq 1-\varepsilon ,
\label{equ:sample_path_bound}
\end{equation}
for all $t \ge 0$. Eq.~\eqref{equ:sample_path_bound} has the form of a complementary cumulative distribution function (CCDF). It is defined, however, for entire sample paths of the process $S(\tau,t)$, i.e, the argument is made for all $\tau \in [0,t]$.
The function $\mathcal{S}^{\varepsilon}(t)$ is known as an $\varepsilon$-effective service curve~\cite{luebben:scest13} that provides a statistical service guarantee with violation probability $\varepsilon$ of the form~\cite{burchard:it06}
\begin{equation}
\mathsf{P} \Bigl[D(t) \ge \inf_{\tau \in [0,t] } \{A(\tau) + \mathcal{S}^{\varepsilon}(t-\tau) \} \Bigr] \geq 1 - \varepsilon ,
\label{equ:stoch_scdefconv}
\end{equation}
for all $t \ge 0$. Effective service curves are a basic system model of the stochastic network calculus~\cite{burchard:it06}. Amongst others, they enable the computation of statistical performance bounds, e.g., a backlog bound $\mathsf{P} \left[ B(t) > b \right] \leq \varepsilon$ follows immediately from $B(t)\!=\!A(t)\!-\!D(t)$ and a delay bound $\mathsf{P} \left[ \Delta(t) > \delta \right] \leq \varepsilon$ from
\begin{equation}
\label{equ:delay_bound}
  \Delta(t)=\inf\{\tau \geq 0: A(t)-D(t+\tau) \leq 0\}
\end{equation}
by use of Eq.~\eqref{equ:stoch_scdefconv}. For illustration, backlog and delay are defined by the vertical and horizontal deviations between the arrival and departure functions, respectively.

To obtain an estimate of $\mathcal{S}^{\varepsilon}(t)$ from measurements, we adapt the method from~\cite{luebben:scest13}. The method uses constant rate probe traffic with rate $r$, i.e., $A(t)\!=\!rt$, and takes measurements of $D(t)$ to obtain the backlog $B(r,t) = rt - D(t)$. For $t \rightarrow \infty$ the steady-state backlog $B(r)$ is observed. Finally, $(1\!-\!\xi)$-quantiles of the steady-state backlog are extracted from
\begin{equation*}
B^\xi(r) = \inf \left\{ x \ge 0: \mathsf{P} \left[ B(r) \le x \right] \geq 1-\xi \right\} ,
\label{equ:w_epsilon}
\end{equation*}
for a set of rates $r \in R$. An estimate of $\mathcal{S}^{\varepsilon}(t)$ is
\begin{equation}
\mathcal{S}^{\varepsilon}(t) = \max_{r \in R} \{ r t - B^{\xi}(r) \} ,
\label{equ:sc_estimate}
\end{equation}
where $\varepsilon\!=\!\sum_{r \in R}{\xi}$ from the union bound. By choice, $\xi$ and the set $R$ are small so that $\varepsilon \ll 1$. An equivalent expression is derived from steady-state delay quantiles $\Delta^\xi(r)$ using the relation $B^\xi(r)\!=\!r\Delta^\xi(r)$ that applies for constant rate traffic and systems that deliver data in order~\cite{luebben:scest13}. In a practical probing scheme, delays can be obtained more easily from send and receive time-stamps of packets than backlogs.

For interpretation of the results, we will frequently use
\begin{equation}
\label{equ:attain_rate}
\alpha^\varepsilon(t)=\frac{\mathcal{S}^{\varepsilon}(t)}{t} ,
\end{equation}
which expresses the service curve as a minimal rate guarantee for a time interval of duration $t$. The rate guarantee may be violated with probability $\varepsilon$. It is related to the notion of available bandwidth that specifies the amount of unused capacity of a link or a network~\cite{luebben:scest13}. For closed-loop flow control, we refer to Eq.~\eqref{equ:attain_rate} as the \emph{attainable rate} function.

An implementation of the method in a practical probing scheme has to estimate the steady-state delay distribution of a TCP connection from measurements of probe traffic sent at different rates. For selection of the probing rates we use linear increments with a defined probing accuracy of $r_{acc}$. The probing starts with rate $r_1\!=\!r_{acc}$. If a steady-state delay distribution can be observed at the current rate, the rate is increased by $r_{acc}$, i.e., the probing rates are $r_j=j \cdot r_{acc}$ with $j=1,2,3,\ldots$. The probing stops at the first rate that does not observe a stationary distribution. This is the case if the arrival rate exceeds the long-term attainable rate, see~\cite{luebben:scest13} for details.

To obtain the steady-state delay distribution $\Delta(r)$ for a probing rate $r$, we randomly sample end-to-end delays from one packet train. After collecting $I$ samples, the samples are tested for stationarity by the Elliot, Rothenberg and Stock test~\cite{Elliott96} and for independence by the runs test~\cite{gibbons2003nonparametric}. If both tests are passed, the samples are used to estimate the steady-state distribution, and the probing proceeds with the next rate. If only stationarity can be assumed~\cite{luebben:scest13}, the probing is repeatedly extended until stationarity and independence are achieved. After the $e$-th extension, $(e+1) I$ samples exist of which each $(e+1)$th sample is selected. The statistical tests are performed on this subset of the samples. Finally, an estimate of the delay quantile $\widetilde{\Delta}^\xi(r)$ is extracted from the empirical distribution of the samples for which stationarity and independence have been detected. A service curve estimate $\widetilde{S}^\varepsilon(t)$ follows from Eq.~\eqref{equ:sc_estimate}, where we use the tilde to indicate estimates that are obtained from empirical data.
\subsection{Evaluation of Window Flow Control}
For a first evaluation, we apply the estimation procedure to basic window flow control protocols. These protocols replicate certain basic functionalities of TCP to investigate them in isolation. Due to the restricted functionality, these protocols are more predictable than the complex TCP.
We present one window flow control protocol with a static window size and one where the window size is adapted to congestion notifications. For the first protocol an analytical service curve can be derived whereas for the second one only the long-term attainable rate is known, e.g., from~\cite{Mathis:tcp}. Both protocols are implemented in the discrete event simulator Omnet++~\cite{omnet}.
\subsubsection{Static Window Flow Control}
\label{sec:static_window_flow_control}
We investigate the flow control loop presented in Fig.~\ref{fig:window_flow_control_scheme} with a static window size $W(t) = w$. The network has a capacity of $C\!=\!10$~Mbit/s and a latency of $T\!=\!50$~ms, i.e., the network service process is $S_{net}(\tau,t)=C[t-\tau-T]^+$. We use a window size of $w=500$~kbit and $w=2$~Mbit, respectively.
\begin{figure}
 \centering
  \subfloat[ window size $w=500$~kbit ]{
    \includegraphics[width=0.47\columnwidth]{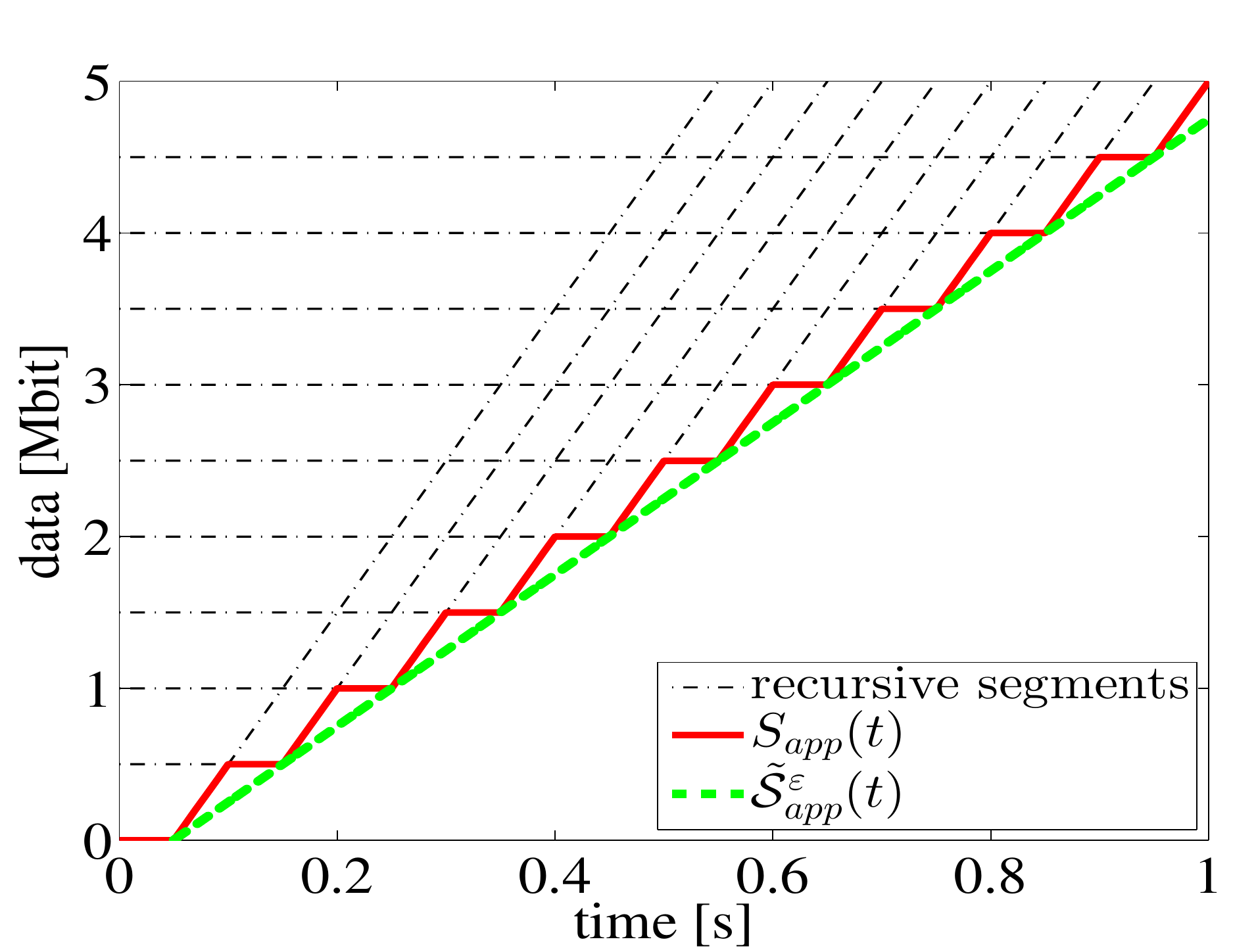}
    \label{fig:flow_control_example}
  }
  \subfloat [ window size $w=2$~Mbit ]{
    \includegraphics[width=0.47\columnwidth]{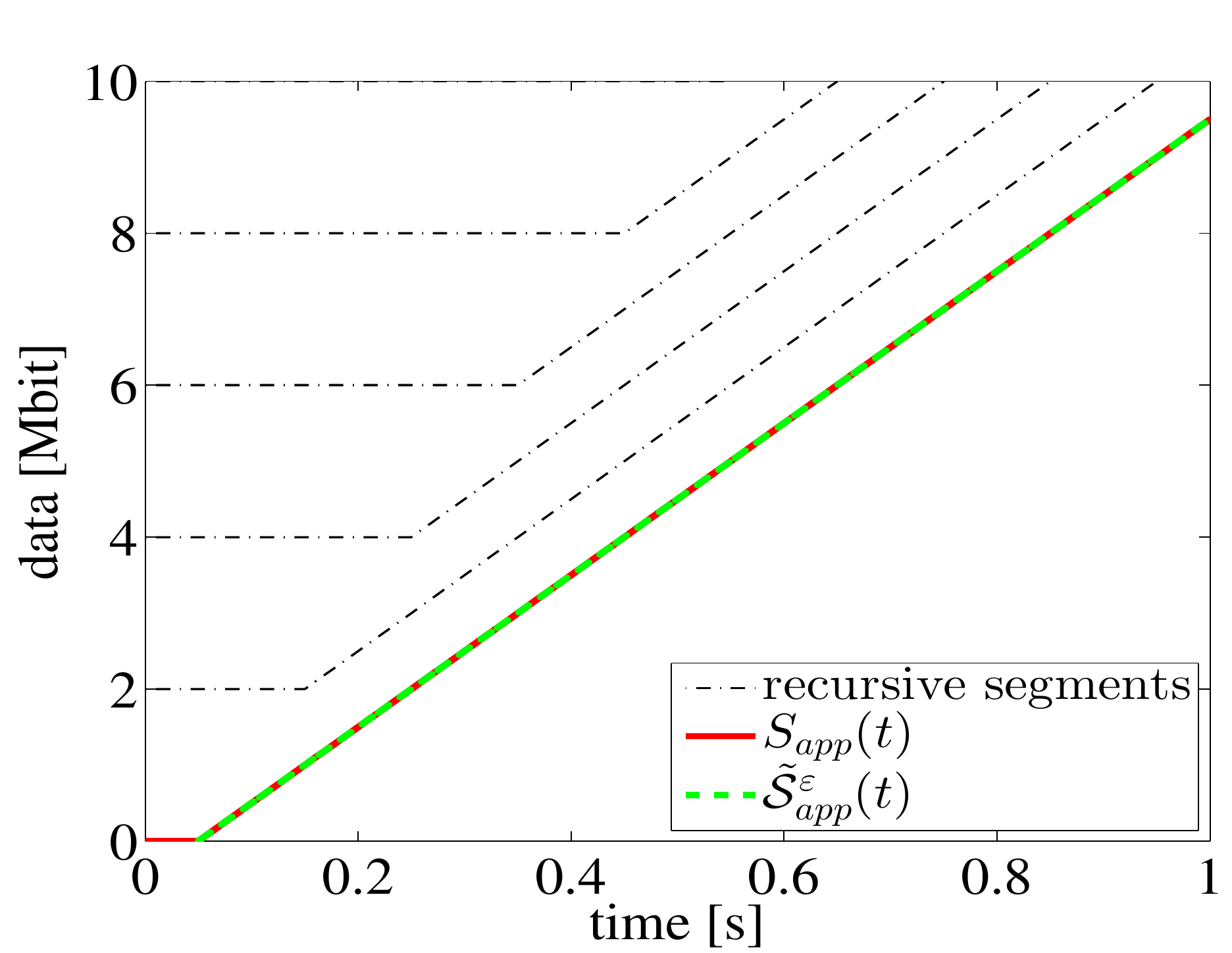}
    \label{fig:flow_control_example_2}
  }
  \caption{Comparison of the analytical service process $S_{app}(t)$ and the estimate $\mathcal{\widetilde{S}}_{app}^{\varepsilon}(t)$. In Fig.~\protect\subref{fig:flow_control_example} the window limits the maximal attainable rate, whereas in Fig.~\protect\subref{fig:flow_control_example_2} the network capacity is fully utilized.}
  \label{fig:flow_control_examples}
\end{figure}

The analytical application level service process $S_{app}(\tau,t)$ follows by recursive insertion of Eq.~\eqref{eq:recursive_fc_a} into Eq.~\eqref{eq:recursive_fc}. The construction is illustrated in Fig.~\ref{fig:flow_control_examples}. $S_{app}(\tau,t)$ is the minimum of infinitely many segments $[r(t\!-\!\tau\!-\!(2n\!+\!1)T)]^+ \! + nw$ for $n=0,1,2,\ldots$ where $r\!=\!\min \{w/RTT, C \}$. Since the system is deterministic, $S_{app}(\tau,t)$ can be expressed as a univariate function $S_{app}(t-\tau)$ that is displayed in Fig.~\ref{fig:flow_control_examples}. The individual segments are indicated by dash-dotted lines. If $w \geq C \cdot RTT$, the service process simplifies to $S_{app}(\tau,t)=C[t-\tau-T]^+$ as shown in Fig.~\ref{fig:flow_control_example_2}. In this case the rate attains the network capacity of 10~Mbit/s. Otherwise, $S_{app}(\tau,t)$ is the stepped curve in Fig.~\ref{fig:flow_control_example}. Here, the window size limits the long-term rate to 5~Mbit/s. In addition, the service process shows in both cases the network latency of 50~ms.

For estimation, we use as accuracy parameter $r_{acc} \!=\! 1$~Mbit/s to obtain an estimate $\mathcal{\widetilde{S}}_{app}^{\varepsilon}(t)$. Given the deterministic system, we let $\varepsilon = 0$, i.e., we use the maximal delay from the empirical delay distribution. The estimate $\mathcal{\widetilde{S}}_{app}^{\varepsilon}(t)$ is presented as dashed line in Fig.~\ref{fig:flow_control_examples}. It is a latency-rate function that is the largest convex lower bound of the analytical result for $w=500$~kbit and exact for $w=2$~Mbit, respectively. We note that Eq.~\eqref{equ:sample_path_bound} defines the effective service curve as a lower bound of the service process. In addition to the long-term rate $\min \{w/RTT, C\}$, the service curves provide information on delays.

As demonstrated here, the method in Sec.~\ref{sec:probe_method} enables the estimation of the service curve for static window flow control. In this setup the delays are apparent from the network structure. For flow control with an adaptive window $W(t)$ no such simple results exist.
\subsubsection{Adaptive Window Flow Control}
\label{sec:basic_window_flow_control}
For the second evaluation, we apply the estimation method to an adaptive window flow control protocol. We again use the flow control loop in Fig.~\ref{fig:window_flow_control_scheme}, with the difference that the window size is adapted dynamically based on random congestion notifications. As a consequence, $S_{app}(\tau,t)$ is random. The protocol replicates the basic congestion avoidance algorithm of TCP.

We simulate a network that does not drop packets due to congestion but instead marks a congestion notification in the packet header. For each packet the receiver sends an acknowledgement that includes the possible congestion notification. We measure the window $w$ in units of packets that are of maximum segment size $MSS$. The window is increased by $1/w$ for each acknowledgement that does not indicate congestion. This leads to a window increase of at most one $MSS$ per $RTT$. In case of a congestion notification, the window size is halved. The control loop emulates the basic congestion avoidance algorithm of TCP NewReno with explicit congestion notification. For this system, the long-term attainable rate is given in~\cite{Mathis:tcp} as $MSS \cdot L / (\sqrt{p} RTT)$, where $p$ is the probability of congestion notifications, and parameter $L\!=\!1.31$ applies for one acknowledgement per packet.

In the experiment, the forward path and the reverse path each have a delay of $5$~ms. A capacity limitation is not simulated. Instead packets are marked with a congestion notification independently of each other with probability $p=10^{-3}$. The long-term attainable rate is calculated as 4143~pkts/s.

\begin{figure}
 \centering
  \subfloat[static vs. adaptive window ]{
    \includegraphics[width=0.47\columnwidth]{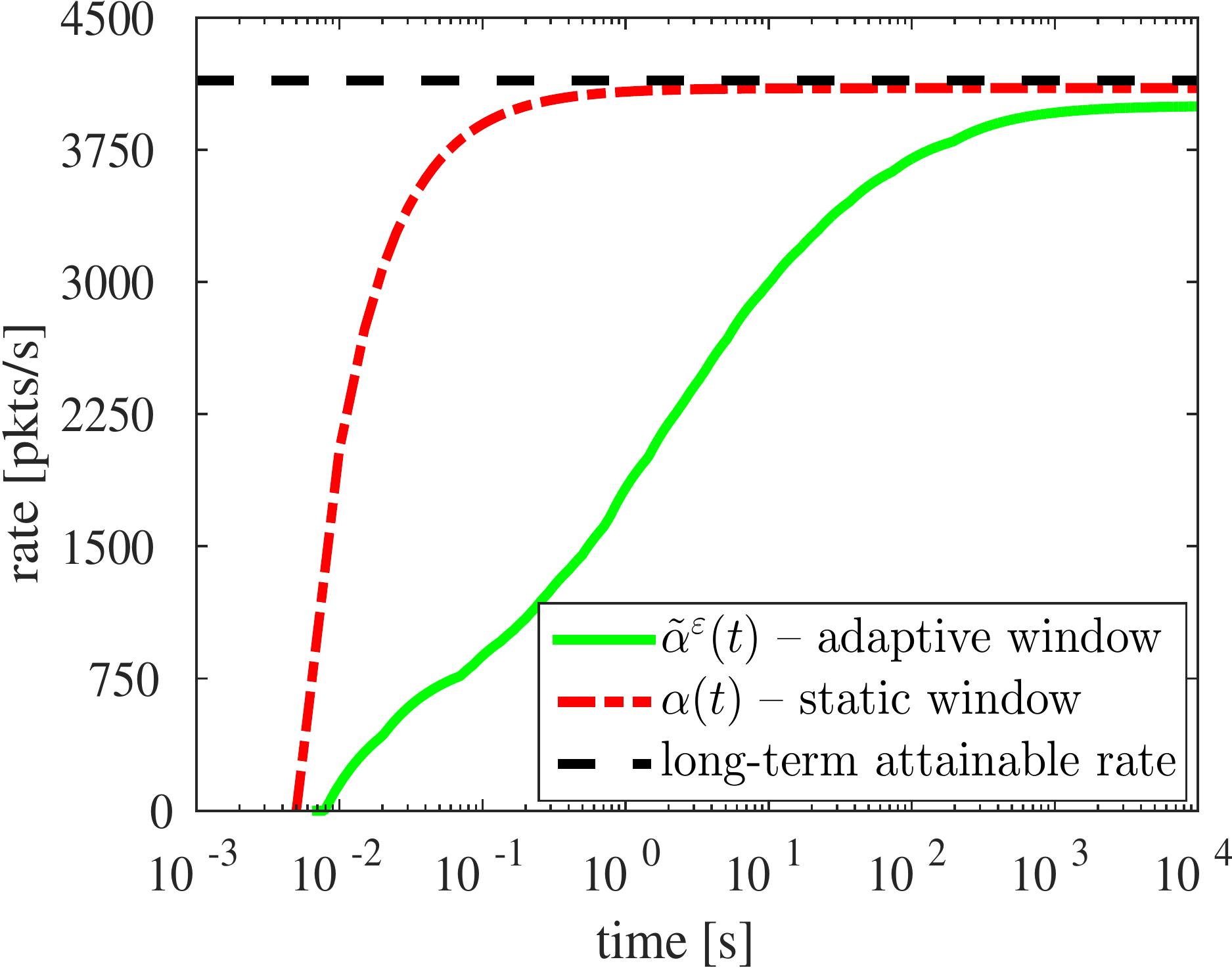}
    \label{fig:adap_flow_control_example}
  }\hfil
  \subfloat [ impact of the accuracy $r_{acc}$]{
    \includegraphics[width=0.47\columnwidth]{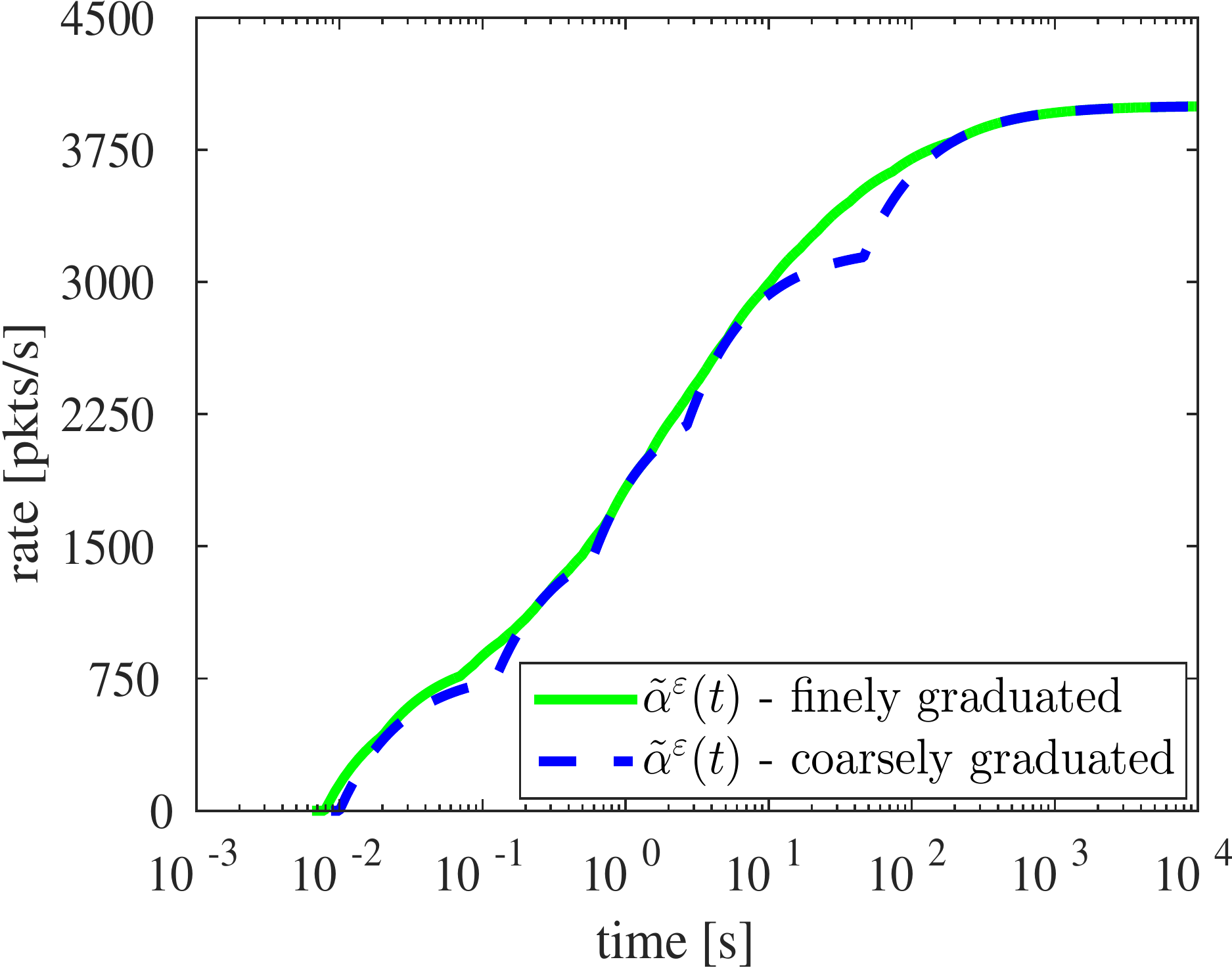}
    \label{fig:adap_flow_control_example_res}
  }
  \caption{Estimates of the attainable rate function $\widetilde{\alpha}^\varepsilon(t)$. Fig.~\protect\subref{fig:adap_flow_control_example} compares flow control protocols with static and adaptive window, respectively. Fig.~\protect\subref{fig:adap_flow_control_example_res} presents the limited resolution if probing rates are coarsely graduated.}
  \label{fig:adap_flow_control_examples}
\end{figure}

We estimate $\mathcal{\widetilde{S}}_{app}^{\varepsilon}(t)$ from delay quantiles, that are obtained using the probing method with $\xi\!=\!10^{-4}$ and $r_{acc}\!=\!100$ \mbox{pkts/s}. Given the target accuracy, the largest probing rate that observes a steady-state is 4100~pkts/s. We present the estimate $\mathcal{\widetilde{S}}_{app}^{\varepsilon}(t)$ as attainable rate function $\widetilde{\alpha}^\varepsilon(t)$ from Eq.~\eqref{equ:attain_rate} labeled adaptive window in Fig.~\ref{fig:adap_flow_control_example}. The attainable rate is zero for $t \!\le\! 5$~ms, i.e., the network delay. For $t \!>\! 5$~ms, it is increasing and converges to 4100~pkts/s, i.e., it recovers the long-term attainable rate of 4143~pkts/s correctly within the granularity of probing.

In Fig.~\ref{fig:adap_flow_control_example}, we also include analytical results of $\alpha^\varepsilon(t)$ from Sec.~\ref{sec:system_model} for flow control with a static window of $w \!=\! 41$~packets, which attains a long-term rate of 4100~pkts/s. The two systems are identical apart from the static, respectively, adaptive window. The comparison shows a strong impact of the adaptive window on the attainable rate. The adaptive window leads to a significantly slower convergence to the long-term rate. Observe that we plot time on log-scale since the deviation is considerable on relevant time-scales from milliseconds up to tens of seconds. The time until the attainable rate of the system with the adaptive window converges is four orders of magnitudes larger than in case of the static window. We note that the effects are not due to an initial transient phase, as we measure the service offered in steady-state.

Simulations enable gathering sufficiently many samples for statistical evaluation. In real networks, this may be difficult due to long observation periods. Therefore, we visualize the effect of the probing accuracy, i.e., parameter $r_{acc}$, on the resolution of the service curve estimate in Fig.~\ref{fig:adap_flow_control_example_res}. Increasing the value is practical for reducing the observation period. The dashed curve shows the estimate for $r_{acc}\!=\!800$~pkts/s compared to the solid line that is obtained for $r_{acc}\!=\!100$~pkts/s. The coarsely graduated rate steps result in a slightly rippled curve $\widetilde{\alpha}^\varepsilon(t)$. In addition, the estimate of the long-term rate has lower resolution, i.e.,~4000~pkts/s, as it is a multiple of $r_{acc}$.
\subsection{Validation of End-to-end Delays}
For an intuitive visualization, we represent the service curve as attainable rate function as defined in Eq.~\eqref{equ:attain_rate} in the previous section. In addition, the service curve property Eq.~\eqref{equ:stoch_scdefconv} enables computing departure guarantees for arbitrary traffic arrivals from $\widetilde{\mathcal{S}}^{\varepsilon}(t)$. In turn, statistical bounds for delay or backlog follow as depicted in the beginning of Sec.~\ref{sec:probe_method}. We make use of this for a numerical validation of delay bounds. For this purpose, we use a traffic trace of an H.264 variable bit rate encoded movie with a mean rate of 2000~pkts/s. This rate corresponds to a utilization of the system described in Sec.~\ref{sec:basic_window_flow_control} of about 0.5. We compare statistical upper bounds of the delay that are obtained from the service curve estimate in Fig.~\ref{fig:adap_flow_control_examples} with empirical delay quantiles observed in simulations for this system.
\begin{figure}[t]
 \centering
  \includegraphics[width=0.8\columnwidth]{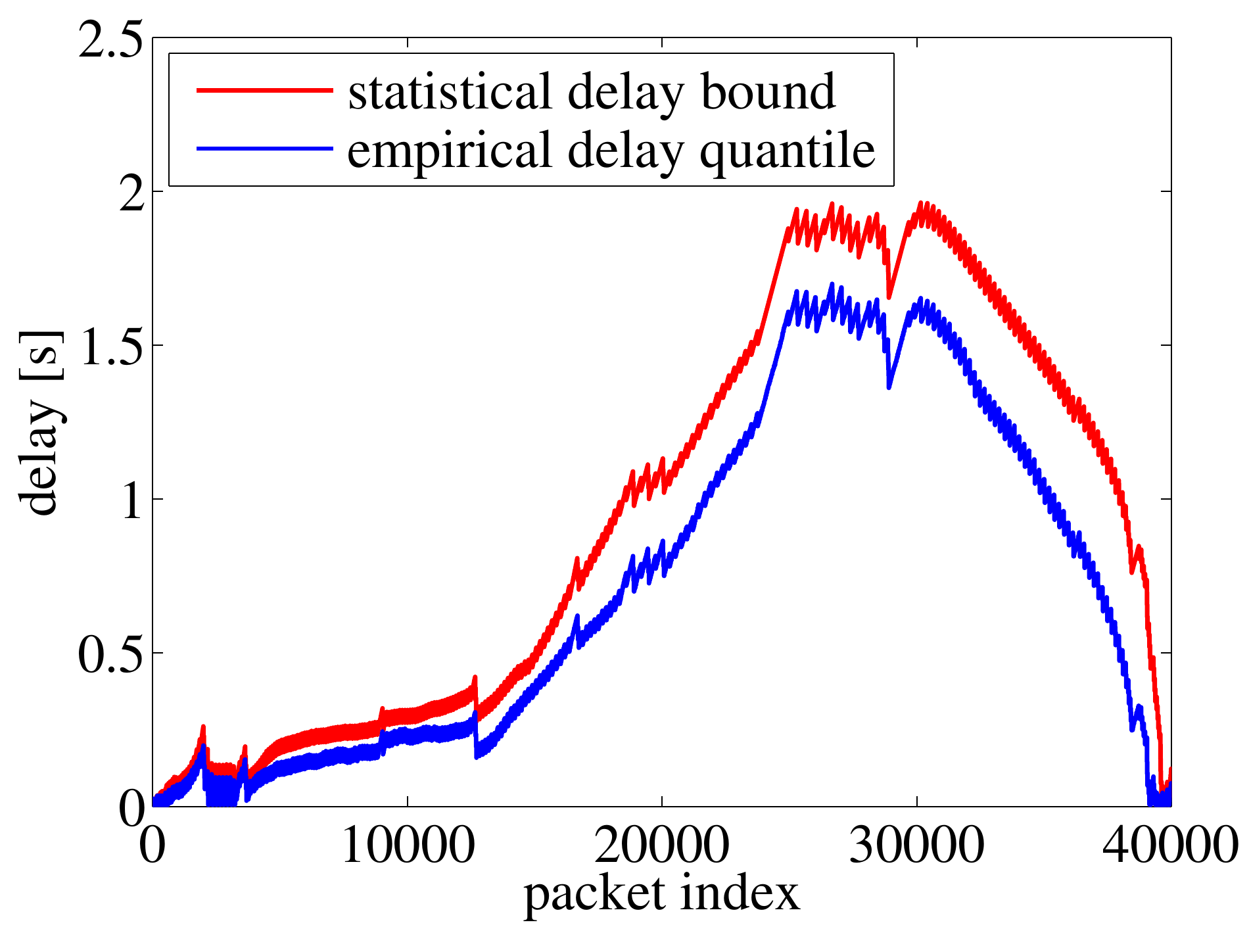}
  \caption{Statistical delay bounds obtained from the service curve estimate compared to empirical delay quantiles obtained from simulations of variable bit rate video traffic. The statistical delay bounds accurately predict the empirical delay quantiles.}
   \label{fig:validation}
\end{figure}

In detail, the statistical delay bounds are computed by Eq.~\eqref{equ:delay_bound} where $A(t)$ are the data arrivals of the video traffic trace. For the data departures $D(t)$ the statistical guarantee provided by Eq.~\eqref{equ:stoch_scdefconv} is employed, where we use the service curve estimate $\widetilde{\mathcal{S}}^{\varepsilon}(t)$ for $\varepsilon = 5 \cdot 10^{-4}$ as obtained in Fig.~\ref{fig:adap_flow_control_example_res}, previously. Note that the statistical delay bound is computed solely from the video arrival traffic trace and the service curve estimate. For comparison, we simulate the transmission of the video traffic by the system. For each packet of the video traffic trace, we empirically obtain $(1\!-\!\varepsilon)$-delay quantiles for $\varepsilon = 5 \cdot 10^{-4}$ from $150000$ repetitions of the simulation.

The statistical upper delay bounds derived from the service curve estimate $\widetilde{\mathcal{S}}^{\varepsilon}(t)$ as well as the corresponding empirically obtained delay quantiles from the simulation are shown in Fig.~\ref{fig:validation} for a series of 40000 packets. Clearly, the service curve estimate provides valid delay bounds that closely follow the series of delay quantiles from simulation. The deviation is a consequence of the granularity of probing $r_{acc}$ and the use of the union bound in Eq.~\eqref{equ:sc_estimate}.
\section{Experimental Setup}
\label{sec:exp_setup}
In Sec.~\ref{sec:basic_window_flow_control}, we implemented a basic flow control loop to illustrate selected mechanisms of a dynamic window adaptation for congestion control as implemented in TCP. Compared to independent congestion notifications, as assumed before, the loss characteristics in networks with aggregated TCP traffic are more complex and a priori unknown. We apply the method from Sec.~\ref{sec:probe_method} in a testbed deployment to demonstrate the applicability to actual protocol implementations and identify effects on TCP's delay performance with respect to various parameters relevant for network dimensioning and application server configuration. The testbed deployment requires, however, more time to collect samples than the simulator. To speed up the estimation, we limit the resolution of the probing method to $r_{acc} = 1$~Mbit/s.

\begin{figure}[t]
\centering
  \includegraphics[width=\columnwidth]{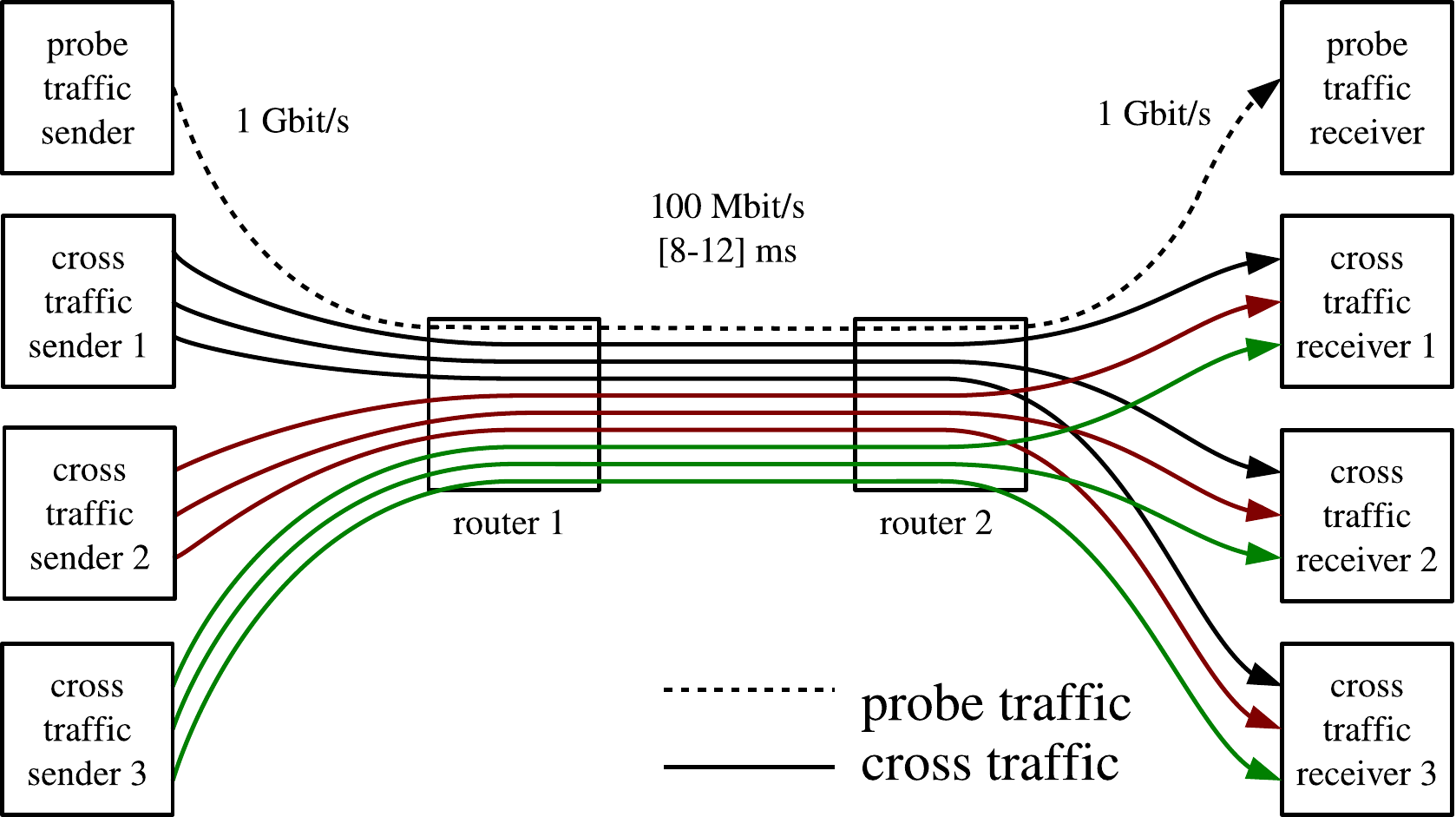}
  \caption[]{Dumbbell network with one probe traffic and nine greedy cross traffic flows. All flows use the same TCP version. Between router 1 and router 2 is a bottleneck link with a capacity of 100 Mbit/s. The end-to-end propagation delay varies for different source destination pairs to prevent synchronization.}
  \label{fig:network_tcp}
\end{figure}

We conducted experiments for various scenarios to investigate the impact of the utilization, buffer size, different AQM schemes, and congestion control protocols. The experiments in the subsequent sections are conducted in a testbed using the Emulab software~\cite{emulab}. All results that we show are obtained for the same network topology, as presented in Fig.~\ref{fig:network_tcp}, to achieve comparability, when we exchange selected characteristic features to evaluate their influence on the performance. The topology is adapted from the common TCP evaluation suite~\cite{tcpevaluationsuite}. Nine greedy TCP NewReno cross traffic flows and one TCP NewReno probe traffic flow, for estimating the attainable rate function $\alpha^\varepsilon(t)$, share the bottleneck link with 100~Mbit/s capacity between router 1 and router 2. The end-to-end propagation delays differ for different source destination pairs to prevent a possible synchronization of the TCP flows. The values are uniformly distributed with a mean of 10~ms as given in Tab.~\ref{tab:one_way_delays_tcp}. The probe traffic flow has a propagation delay of 10~ms. Since TCP is designed to achieve fairness between flows that share the same bottleneck link, the expected long-term attainable rate of a TCP connection is 10~Mbit/s gross, respectively, 9.4~Mbit/s net at the application layer.
\begin{table}[t]
\caption{Propagation delays of network paths.}
\begin{center}
  \begin{tabular}{|c|c|c|c|c|}
  \hline
 			& probe  	& cross 	& cross 	& cross	 \\
	delay [ms]	& traffic  	& traffic  	& traffic 	& traffic \\
			& receiver	& receiver 1  	& receiver 2	& receiver 3\\ \hline
probe traffic sender	&	10		& -		& - 	& -	\\ \hline
cross traffic sender 1 	&	-		& 12		& 11 	& 9	\\ \hline
cross traffic sender 2	&	-		& 11		& 8 	& 10	\\ \hline
cross traffic sender 3	&	-		& 9		& 8 	& 12	\\ \hline
\end{tabular}
\end{center}
\label{tab:one_way_delays_tcp}
\end{table}

All nodes run the Linux operating system with kernel 3.5, i.e., all experiments are based on the network stack implementation available for that kernel. Following the guidelines from~\cite{bestpractice}, offloading features on all network interface cards (NICs) are disabled. The hierarchical token bucket filter is used for rate limiting to avoid effects by non-controllable buffers in drivers or NICs. Socket buffers at the sender and the receiver sides are adjusted, so that these do not limit the performance of the connection. The clocks of all nodes are synchronized.

If not stated otherwise, the following parameters are used by the method from Sec.~\ref{sec:probe_method} to estimate the attainable rate function $\alpha^\varepsilon(t)$. We compute $0.95$-quantiles from empirical steady-state delay distributions. For each distribution 750 statistically independent samples are captured. The packet size is 1500~Byte.
\section{Evaluation of TCP NewReno}
\label{sec:eval_reno}
We begin this section with an evaluation of the attainable rate function of TCP NewReno by using the method from Sec.~\ref{sec:probe_method}. The results illustrate the complete behavior of a TCP connection including the dynamics of the protocol stack. Using a decomposition of end-to-end delays as in Fig.~\ref{fig:abstraction_network_path}, this enables the identification of effects that have a strong impact on the attainable rate. Moreover, we illustrate the root cause for large end-to-end delays. For all experiments, we used the topology introduced in Sec.~\ref{sec:exp_setup}. Focusing on the basic topology and TCP NewReno enables us to connect the estimation results to essential characteristics of TCP's additive increase multiplicative decrease (AIMD) congestion control.
\subsection{Attainable Rate Function}
We estimate the attainable rate function for a network in which FIFO drop-tail queueing is configured and the buffer sizes are set to the BDP of 100~Mbit/s times 20~ms. Using packets of 1500~Byte size, the buffers can hold 167~packets.
\begin{figure}[t!]
 \centering
 \includegraphics[width=0.8\columnwidth]{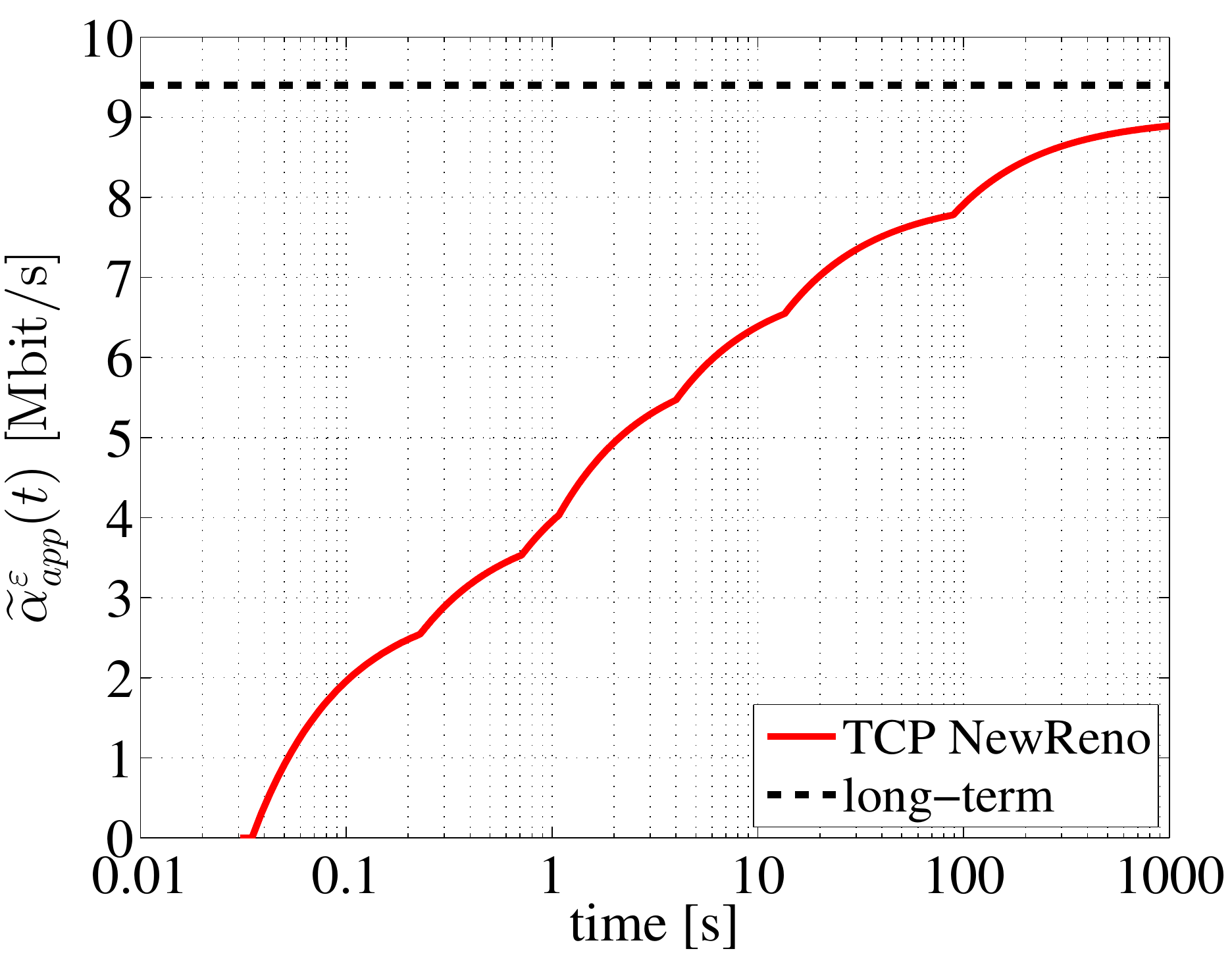}
  \caption{Estimate of the attainable rate function for TCP NewReno. On short time-scales the rate is substantially below the long-term rate. It converges on long time-scales.}
  \label{fig:tcp_attainable_rate_newreno}
\end{figure}
We show the estimate of the attainable rate function $\widetilde{\alpha}^\varepsilon(t)$ on logarithmic time scale in Fig.~\ref{fig:tcp_attainable_rate_newreno}. The origin of the attainable rate is at 30~ms, which corresponds to the propagation delay of 10~ms plus the maximal queueing delay at the bottleneck link of 20~ms. The long-term attainable rate converges to 9~Mbit/s, that is the fair share of the bottleneck capacity for a single TCP flow, within the accuracy of the probing of $r_{acc}\!=\!1$~Mbit/s. The method correctly identifies the characteristics that are specific to the topology. Beyond that, the method reveals the slow speed of convergence to the long-term attainable rate, which is remarkable. On relevant time scales up to several seconds, the attainable rate is significantly below the fair share, i.e., applications have to anticipate rates significantly below the long-term attainable rate.
\subsection{Decomposition of End-to-end Delays}
\label{sec:decomposition}
To understand the underlying effects, we explore the causes of large end-to-end delays. Recall that the attainable rate function is constructed from linear rate segments and corresponding delay quantiles, where larger delays give rise to a slower convergence. The delay quantiles are extracted from the distribution of the end-to-end steady-state delays of a TCP connection denoted $\Delta_{e2e}$. We decompose $\Delta_{e2e}$ into the delay in the sender's protocol stack $\Delta_{snd}$, in the network $\Delta_{net}$, and in the receiver's protocol stack $\Delta_{rcv}$, see Fig.~\ref{fig:abstraction_network_path}. This decomposition identifies which segments contribute most to the end-to-end delay. The results are presented as CCDF of the stationary delay distribution for a moderate arrival rate. The presentation as CCDF covers the complete characteristics instead of an extract as the time-series presented in Fig.~\ref{fig:cp_queue_size_delay}.

To obtain the delays, we measure the following packet time-stamps: $t_1$ creation of a packet at the sender's application, $t_2$ transfer from the kernel to the NIC at the sender, $t_3$ transfer from the NIC to the kernel at the receiver, and $t_4$ reception of the packet at the receiver's application. The delays are computed as $\Delta_{e2e}\!=\!t_4\!-\!t_1$, $\Delta_{snd}\!=\!t_2\!-\!t_1$, $\Delta_{net}\!=\!t_3\!-\!t_2$, and $\Delta_{rcv}\!=\!t_4\!-\!t_3$. The transfer time between NIC and kernel is captured with libpcap.

\begin{figure}
    \centering
    \includegraphics[width=0.8\columnwidth]{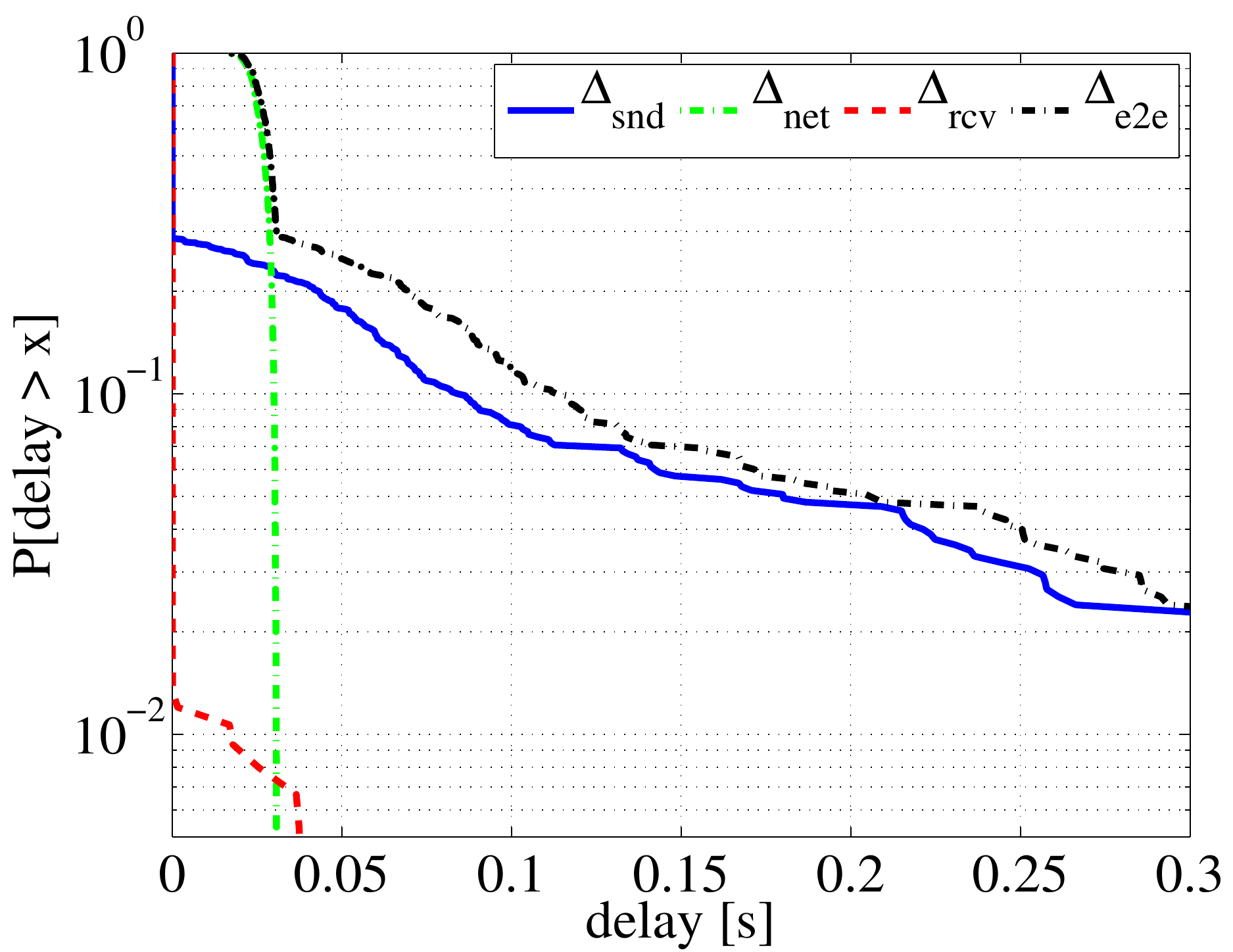}
    \caption{CCDFs of steady-state delays for an application layer rate of 5~Mbit/s. The end-to-end delay $\Delta_{e2e}$ is decomposed into the delay in the sender's protocol stack $\Delta_{snd}$, the network $\Delta_{net}$, and the receiver's protocol stack $\Delta_{rcv}$. The results show a strong impact of $\Delta_{snd}$ on $\Delta_{e2e}$.}
    \label{fig:delay_segment}
\end{figure}

The individual delay components for a moderate arrival rate of 5~Mbit/s are presented in Fig.~\ref{fig:delay_segment}. The network delay $\Delta_{net}$ ranges from 18~ms to 30~ms, whereof the propagation delay accounts for a constant amount of 10~ms. The transmission and processing delays are negligible. Hence, after subtraction of the propagation delay, a queueing delay between 8~ms and 20~ms remains, where 20~ms corresponds to the maximal queueing delay at the BDP-sized buffer. The CCDF of $\Delta_{net}$ shows that none of the sampled packets observed an empty bottleneck buffer and that large queueing delays close to the maximum occur frequently. The effect is due to the high utilization of the bottleneck link that is caused by the greedy TCP cross traffic flows.

In the sender's protocol stack, delays arise that are more than ten times larger than the network delay, as shown by the CCDF of $\Delta_{snd}$. The probability that a packet is queued in the stack is about 0.3, however, the decay of the tail distribution is much slower than for the network delay. While the queueing delays in the network exhibit a faster than exponential decay, the delay distribution in the sender's stack has approximately an exponential decay, observable by the linear decay of the CCDF on the logarithmic scale. We explore the exponential decay in detail in~\cite{luebben:TCPdelaydist}. We note that the delays in the sender's protocol stack are caused by the adaptive congestion window only, as we configured sufficiently large send and receive buffers that do not limit the service. We observed a similar impact of the congestion window in the simulation results presented in Sec.~\ref{sec:basic_window_flow_control}.

Delays at the receiver's protocol stack $\Delta_{rcv}$ occur only with a small probability. They are caused by buffering of packets that are received after a packet is lost, to guarantee in-order delivery of the packets.

Concluding, we highlight that the significant delay portions explained before are clearly observable from the end-to-end delay $\Delta_{e2e}$. The CCDF of $\Delta_{e2e}$ exhibits two regions that are due to the network and the sender's protocol stack, respectively, separated by a sharp bend at the probability at which delays occur in the sender's protocol stack.

\begin{figure}
    \centering
    \includegraphics[width=0.8\columnwidth]{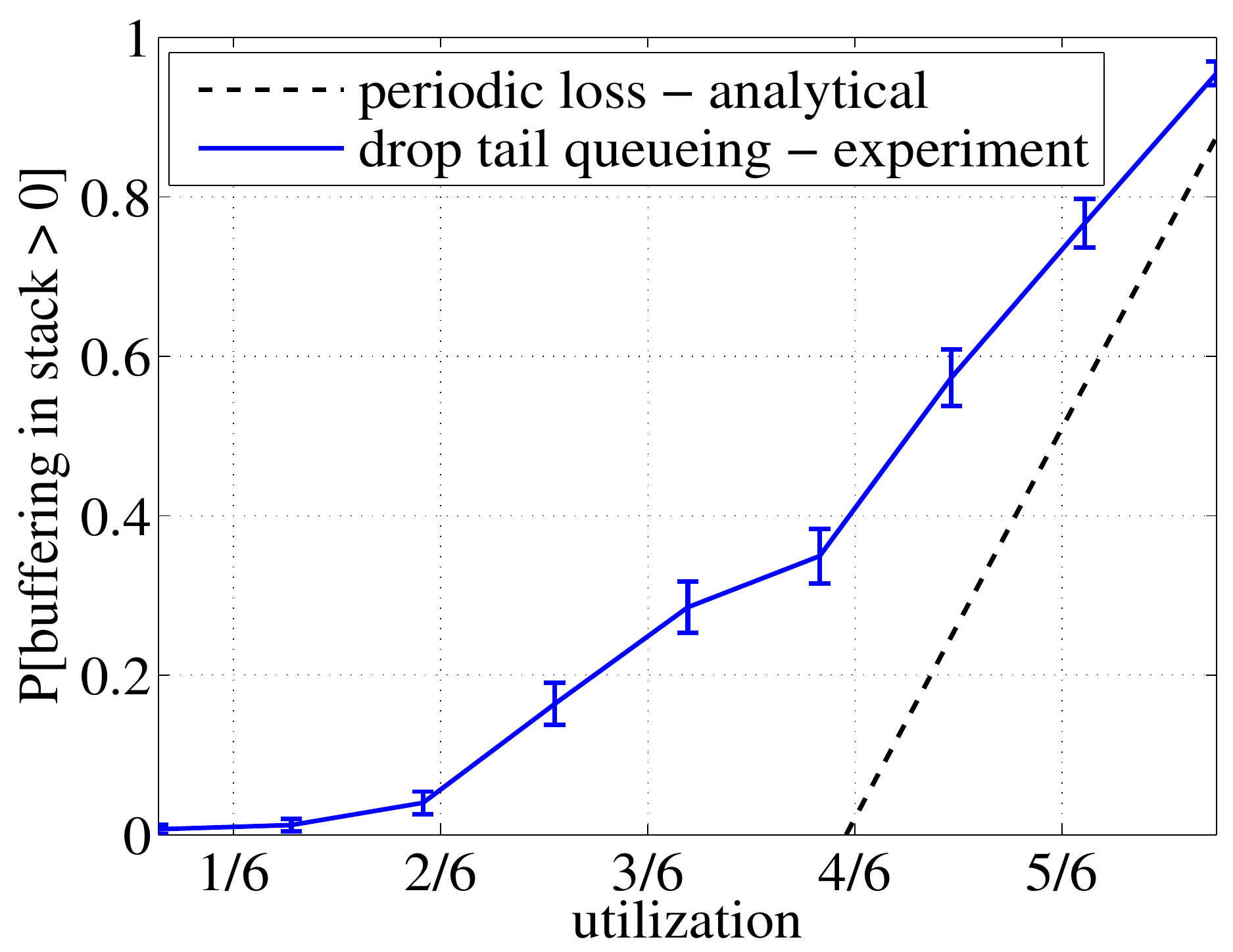}
    \caption{Probability of buffering in the sender's protocol stack for periodic and random losses, respectively.}
    \label{fig:prob_stack_queueing}
\end{figure}
\subsection{Root Cause for End-to-end Delays}
As we observed, significant delays occur already for moderate arrival rates. We explain the accumulation of such high delays in the sender's protocol stack. For illustration, we first consider the basic TCP throughput model for periodic losses with probability $p$ as in~\cite{Mathis:tcp}. Under the assumptions of this model, the CWND $W(t)$ oscillates between $W_{max}\!=\!\sqrt{8/(3p)}$ and $W_{\min}\!=\!W_{\max}/2$. The oscillation results from the AIMD congestion avoidance algorithm of TCP NewReno. It follows that the instantaneous transmission rate, that is given by $W(t)/RTT$, does not fall below $W_{\max}/(2 RTT)$. The long-term attainable rate is $3 W_{\max}/(4 RTT)$ as determined from the average window size $3 W_{\max}/4$. Hence, the minimal instantaneous transmission rate of $W_{\max}/(2 RTT)$ corresponds to a utilization of the long-term rate of 2/3. Consequently, delays in the sender's protocol stack should not occur as long as the application generates data at a rate below 2/3 of the long-term attainable rate of the TCP connection. Obviously, as presented before, delays in the sender's protocol stack occur, however, for much smaller application data rates.

We display the probability of buffering in the sender's protocol stack as a function of the utilization in Fig.~\ref{fig:prob_stack_queueing}. The utilization is measured as the quotient of the arrival rate and the long-term attainable rate, i.e., the fair share of 9.4~Mbit/s of the TCP connection. Already for small utilizations buffering in the sender's protocol stack occurs. If the utilization approaches one, the probability of buffering goes to one. The behavior for small utilizations is notable if compared to the basic TCP throughput model with periodic losses~\cite{Mathis:tcp}. This strong discrepancy between the periodic loss model and the experimental results stems from the adaptation of the congestion window and non-periodic, random losses, which occur for drop-tail queueing.

\begin{figure}
  \centering
    \includegraphics[width=0.8\columnwidth]{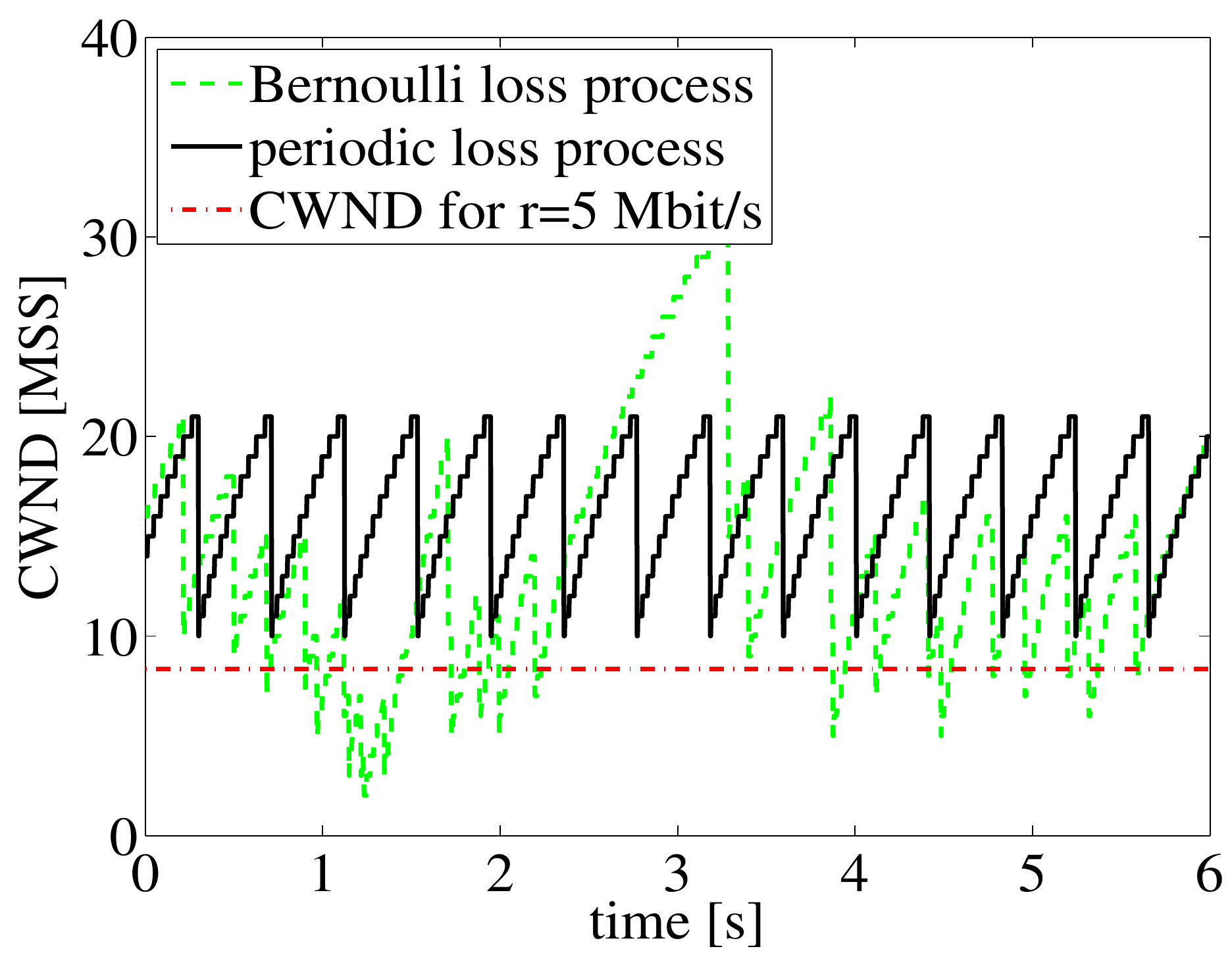}
  \caption{CWND series for Bernoulli and periodic losses, respectively. In case of Bernoulli losses, small CWND values below the CWND that is required to achieve a rate of 5 Mbit/s sustain for durations in the range of hundreds of milliseconds and account for large delays in the sender's protocol stack.}
  \label{fig:cwnd_series_synthetic}
\end{figure}
We investigate the accumulation of delays in the sender's protocol stack by simulation to illustrate the causal effects. The simulation parameters are adjusted to the experimental testbed with a long-term rate of 9.4~Mbits/s and an RTT of 20~ms. The basic congestion avoidance algorithm as assumed in~\cite{Mathis:tcp} and outlined in Sec.~\ref{sec:basic_window_flow_control} is implemented for adaptation of the CWND. The application data rate is set to 5~Mbit/s corresponding to a moderate utilization. Fig.~\ref{fig:cwnd_series_synthetic} demonstrates the reason for delays by comparing a CWND series for random losses due to a Bernoulli process to periodic losses. Random packet losses may be clustered such that the CWND is forced below the value of $W_{min}$ as computed for periodic losses. Further, the CWND that is required to support the application layer data rate of 5~Mbit/s is indicated in Fig.~\ref{fig:cwnd_series_synthetic} by a dash-dotted horizontal line. For random losses, the instantaneous CWND falls repeatedly below this value so that considerable buffering in the sender's stack occurs, whereas in case of periodic losses it generally stays above. While we illustrated the origin of large delays using a simplified simulation model that is based on~\cite{Mathis:tcp}, we note that further effects occur in real-world implementations that have an impact on delays, such as delayed acknowledgements, time-varying RTTs, clustered losses, and implementations of the congestion control algorithm in operating systems. We benchmark these effects next.

\begin{figure*}
 \centering
  \subfloat[ rate 1 to 3 Mbit/s ]{
    \includegraphics[width=0.32\textwidth]{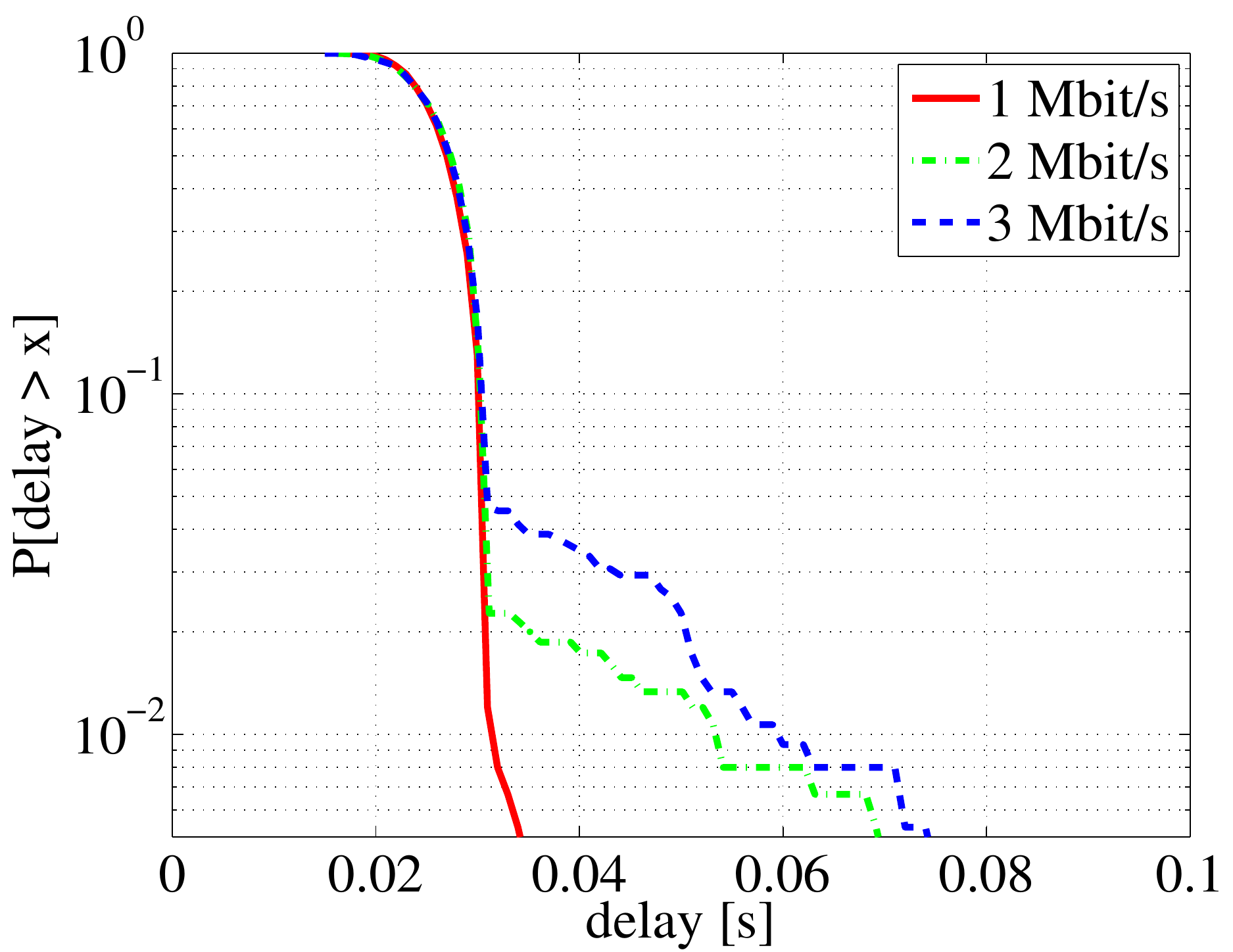}
    \label{fig:delay_ccdf_reno_167_1fit2}
  }
  \subfloat [ rate 4 to 6 Mbit/s ]{
    \includegraphics[width=0.32\textwidth]{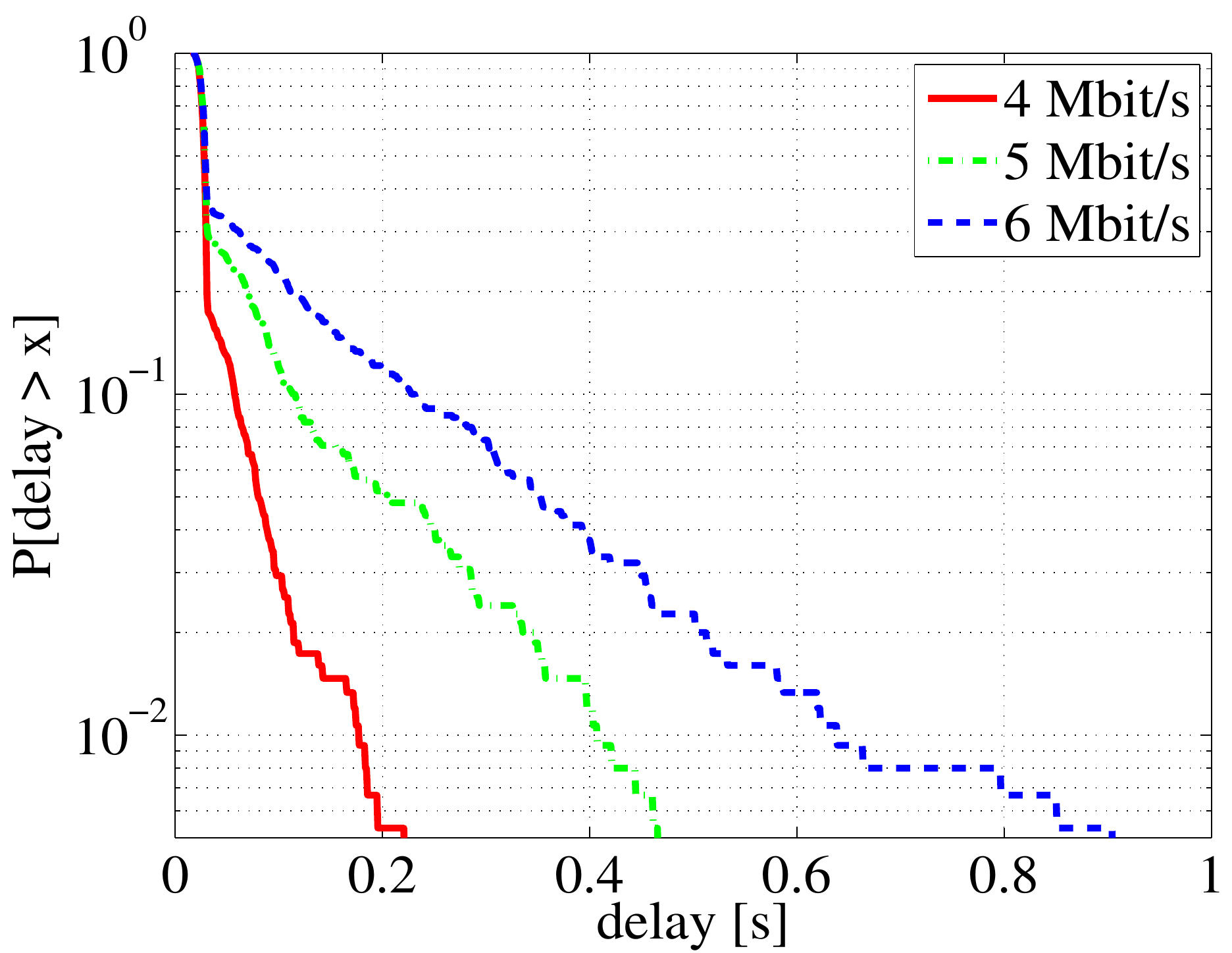}
    \label{fig:delay_ccdf_reno_167_5fit2}
  }
   \subfloat [ rate 7 to 9 Mbit/s ]{
    \includegraphics[width=0.32\textwidth]{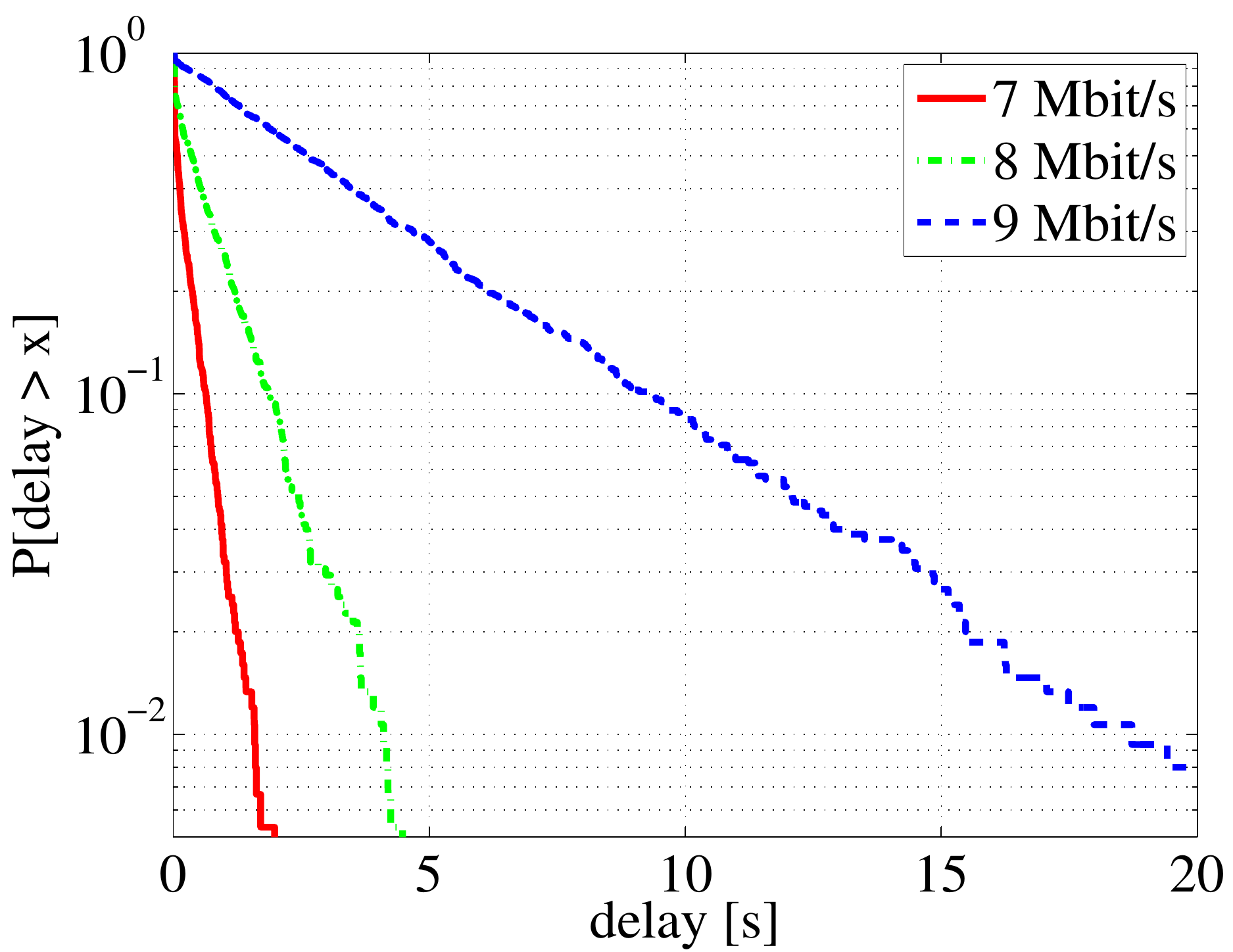}
    \label{fig:delay_ccdf_reno_167_9fit2}
  }
  \caption{CCDFs of end-to-end delays for probing rates from 1~Mbit/s to 9~Mbit/s. All CCDFs exhibit two regions: one region up to about 30~ms originating from queueing in the network, that decays faster than exponentially, and a second region starting from about 30~ms caused by buffering in the sender's protocol stack, that decays approximately exponentially fast.}
  \label{fig:delay_ccdf_reno_fit_rates}
  \vspace{-1em}
\end{figure*}
\section{Benchmark}
\label{sec:benchmark2}
Next, we use estimates of the attainable rate function to benchmark essential configuration options of TCP connections. We start with the impact of the utilization, as typically evaluated in queueing theory. Furthermore, we benchmark the buffer size for drop-tail queueing, which is often deployed in computer networks, in Sec.~\ref{sec:drop-tail}. Thereby, packet loss occurs if buffers overflow, which is directly related to the queue size. To avoid large queues and burst losses AQM schemes were developed. We evaluate different AQM schemes in Sec.~\ref{sec:aqm}. The rate adaptation due to loss is implemented by congestion control protocols, which we compare in Sec.~\ref{sec:congestion_control}. Lastly, we investigate differences in the performance for these options. For all experiments, we use the topology introduced in Sec.~\ref{sec:exp_setup}.
\subsection{Impact of the Utilization}
The CCDF of the end-to-end delays exhibits two distinctive regions, one due to network delays, and one due to delays in the sender's protocol stack. We examine the impact of the utilization on these regions for different probing rates from 1~Mbit/s to 9~Mbit/s with increments of 1~Mbit/s. With respect to the fair share of a flow, the rates correspond to utilizations of about 0.1 to 0.9. The CCDFs of the end-to-end delay are depicted in Fig.~\ref{fig:delay_ccdf_reno_fit_rates}. We discover that all CCDFs show the two distinctive regions with a similar decay for all rates: the first region decays faster than exponentially whereas the second region exhibits an approximately exponential decay. The transition of the first to the second region is at about 30~ms, which equals the maximum network delay. In~\cite{luebben:TCPdelaydist}, we further analyze the delay distribution and confirm its exponential decay.

Using the decomposition of delays as in Fig.~\ref{fig:delay_segment}, the two regions arise from delays in the network and delays in the sender's protocol stack, respectively. As Fig.~\ref{fig:delay_ccdf_reno_fit_rates} shows, the intensity and the probability of delays due to buffering in the sender's protocol stack grow substantially with increasing arrival rate. For low rates, Fig.~\ref{fig:delay_ccdf_reno_167_1fit2}, the probability that end-to-end delays above the maximum network delay occur is small and their intensity is low. In contrast, for high rates, Fig.~\ref{fig:delay_ccdf_reno_167_9fit2}, the probability of delays due to buffering in the sender's stack is close to one and delays of several seconds occur with non-negligible probability. Remarkably, even for moderate rates much below the fair share, Fig.~\ref{fig:delay_ccdf_reno_167_5fit2}, significant delays occur due to buffering in the sender's protocol stack. These delays can be magnitudes above the maximum network delay. Consequently, a high utilization has a significant impact on the delay performance, where we identify a more than proportional relation, see~\cite{luebben:TCPdelaydist}. Reducing the utilization, e.g., by over-provisioning or rate adaptation at the application, can reduce the delay effectively.
\subsection{Impact of the Buffer Size}
\label{sec:drop-tail}
In network dimensioning, a conventional rule of thumb states that the buffer size at routers should be set to the BDP. The rule is justified in a scenario where a single TCP NewReno flow saturates a link. It is demonstrated that the buffer size and thereby the queueing delay can be reduced significantly without compromising the link utilization, e.g., if many flows are present~\cite{Appenzeller:2004,tinyBuffers:2008} or by supporting two queues of different size~\cite{Podlesny:rateDelayTradeoff}, a small queue for delay sensitive flows and a large one for bulk data transfers.

The previous experiment showed that end-to-end delays may be larger than queueing delays in router buffers by orders of magnitude. This raises the question, how the size of router buffers influences end-to-end delays? The estimation of the attainable rate $\widetilde{\alpha}^\varepsilon(t)$ is applicable to provide a measure of these effects. Compared to~\cite{Appenzeller:2004,tinyBuffers:2008}, where the link utilization for aggregated TCP traffic is analyzed, the attainable rate $\widetilde{\alpha}^\varepsilon(t)$ examines the performance of individual end-to-end connections.

We perform the estimation for buffer sizes in the range from 1/8 up to 3/2 times the BDP of 167~packets configured on the interfaces between the two routers. We show results for selected buffer sizes of 21, 83, 167, and 251 packets, respectively. Further results are omitted to keep the presentation manageable.
\subsubsection{Smaller Buffers -- Faster Convergence}
\begin{figure}
 \centering
  \subfloat[ short-term ]{
    \includegraphics[width=0.47\columnwidth]{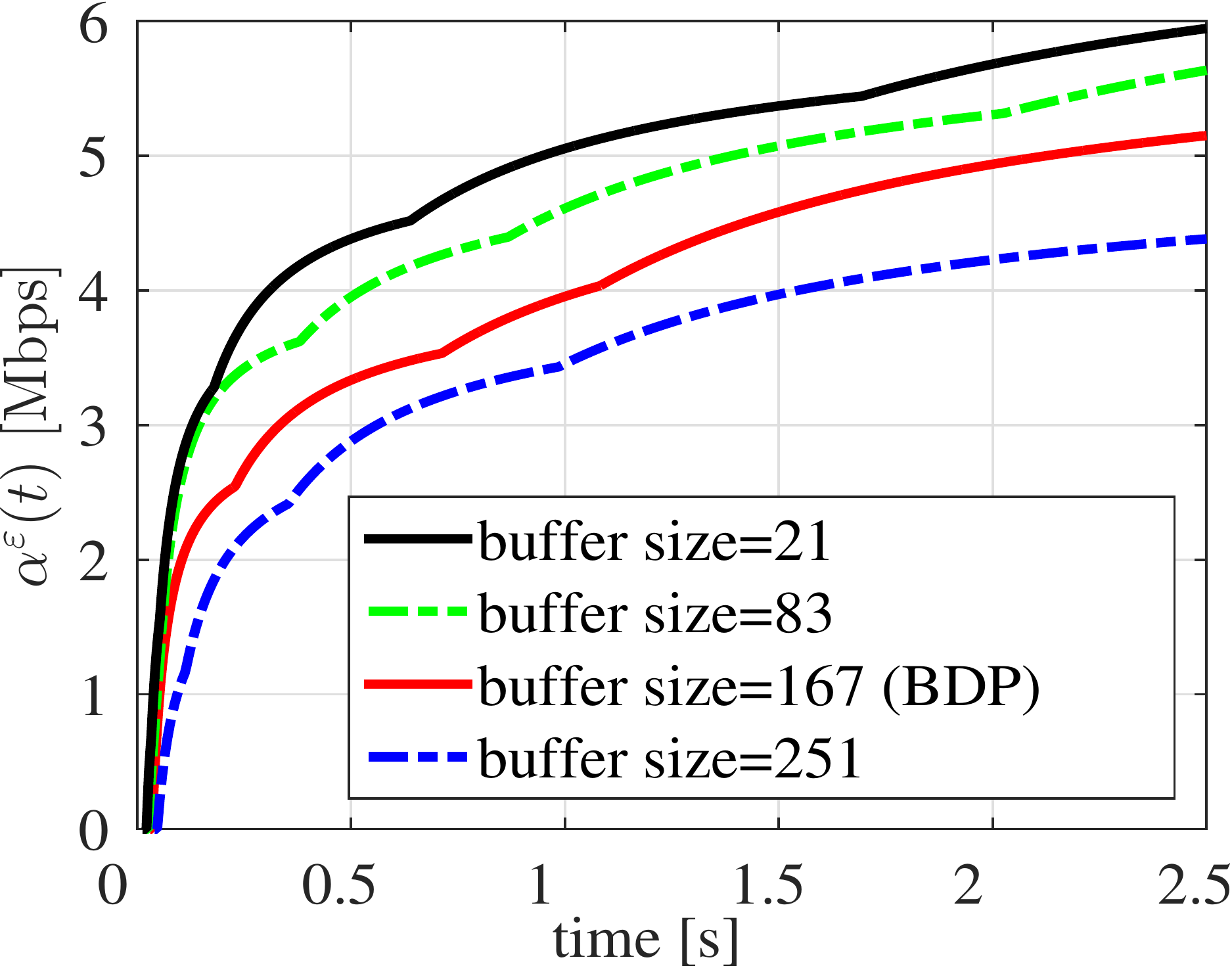}
    \label{fig:tcp_queue_size_st}
  }
  \subfloat [ long-term ]{
    \includegraphics[width=0.47\columnwidth]{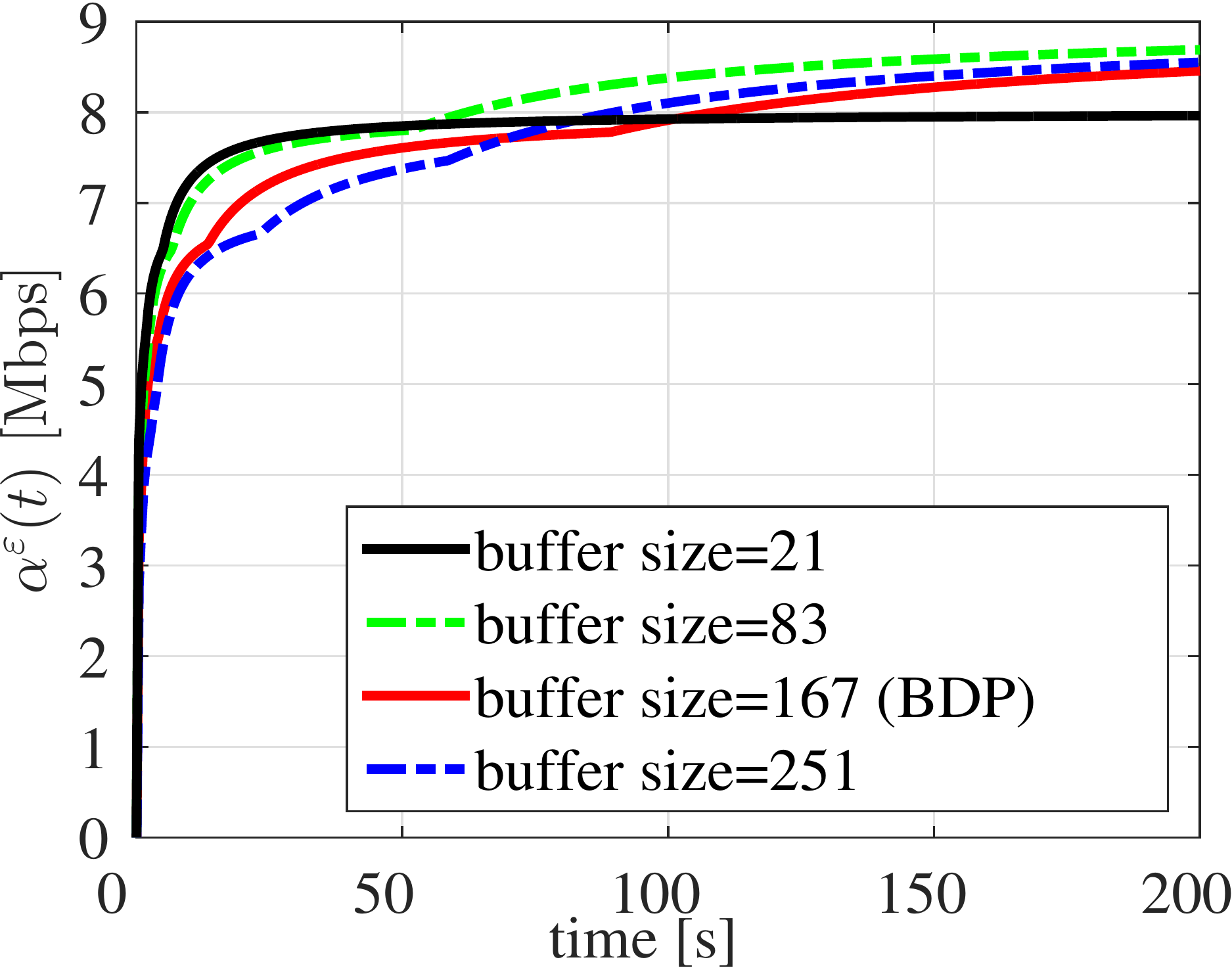}
    \label{fig:tcp_queue_size_lt}
  }
  \caption{Estimates of the attainable rate function for various buffer size configurations. Small buffers, clearly below the BDP, increase the attainable rate on short time-scales. Undersized buffers, however, limit the long-term attainable rate below the fair share of the TCP connection.}
  \label{fig:tcp_queue_size}
\end{figure}

In Fig.~\ref{fig:tcp_queue_size}, we depict the attainable rate function $\widetilde{\alpha}^\varepsilon(t)$ of a TCP NewReno connection for different router buffer sizes. Ideally, $\widetilde{\alpha}^\varepsilon(t)$ attains the fair share of the TCP connection that is 9.4~Mbit/s net. For all tested buffer sizes above 21 packets, Fig.~\ref{fig:tcp_queue_size_lt} shows that $\widetilde{\alpha}^\varepsilon(t)$ converges to 9~Mbit/s, i.e., the fair share is achieved within the configured probing accuracy of $r_{acc}\!=\!1$~Mbit/s. Only for a very small buffer size of 21~packets, the long-term attainable rate is limited to 8~Mbit/s. The result extends the finding that aggregate TCP traffic saturates a link, even if the buffer size is much smaller than the BDP~\cite{Appenzeller:2004,tinyBuffers:2008}. In addition our result also includes the aspect of fairness, as it shows that individual TCP connections can utilize their fair share of the link capacity.

While small buffers affect the long-term attainable rate only marginally, they improve the attainable rate on short time-scales considerably. As illustrated in Fig.~\ref{fig:tcp_queue_size_st}, small router buffers achieve a significantly faster convergence of the attainable rate function to the long-term rate. For example on a time-scale of one second the attainable rate improves by up to 50\%.
\subsubsection{Delays and the Speed of AIMD}
\label{sec:drop-tail-speed}
\begin{figure}
 \centering
 \subfloat[ CCDFs ]{
    \includegraphics[width=0.47\columnwidth]{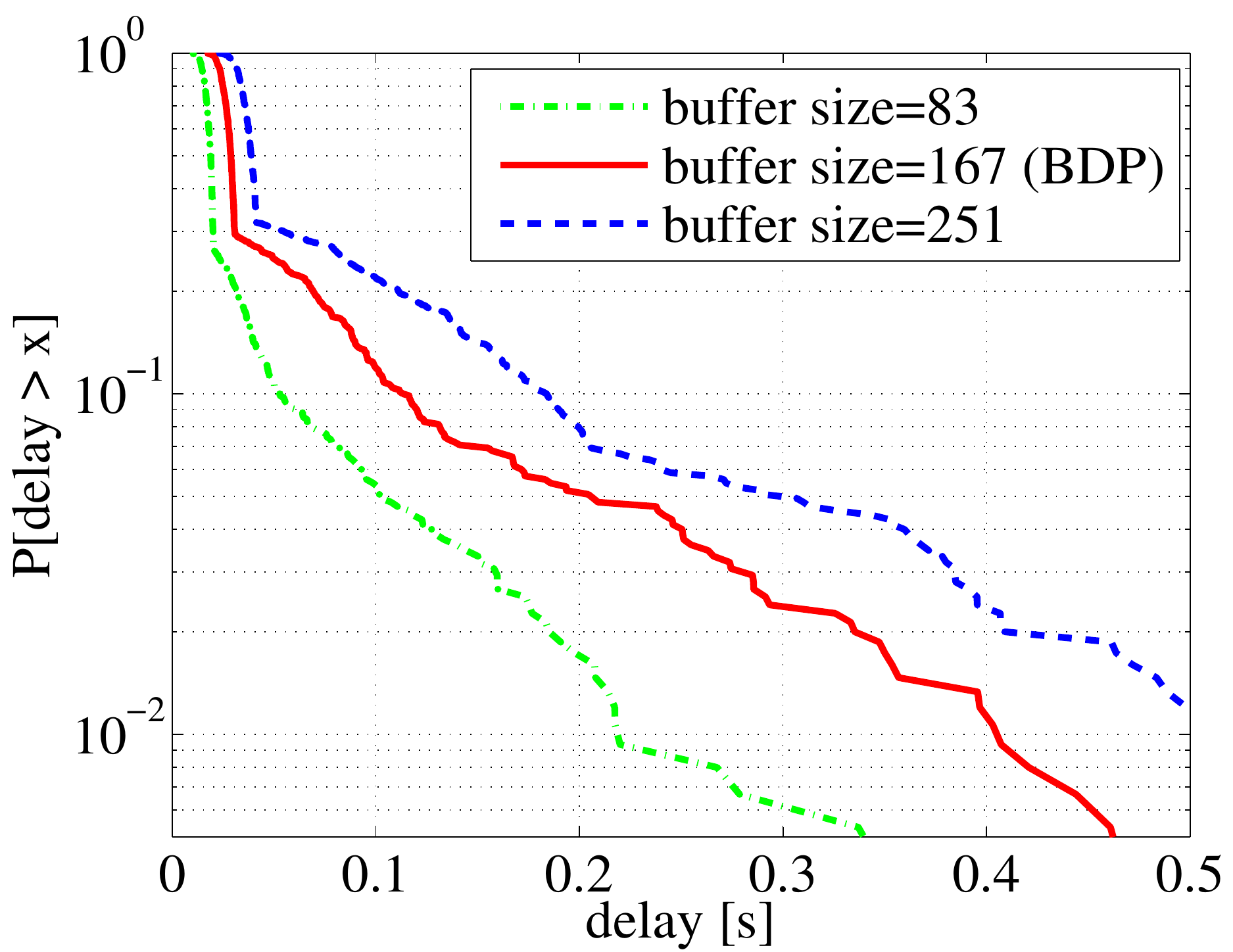}
    \label{fig:ccdf_queue_size}
  }
  \subfloat [ autocorrelation of the CWND ]{
    \includegraphics[width=0.47\columnwidth]{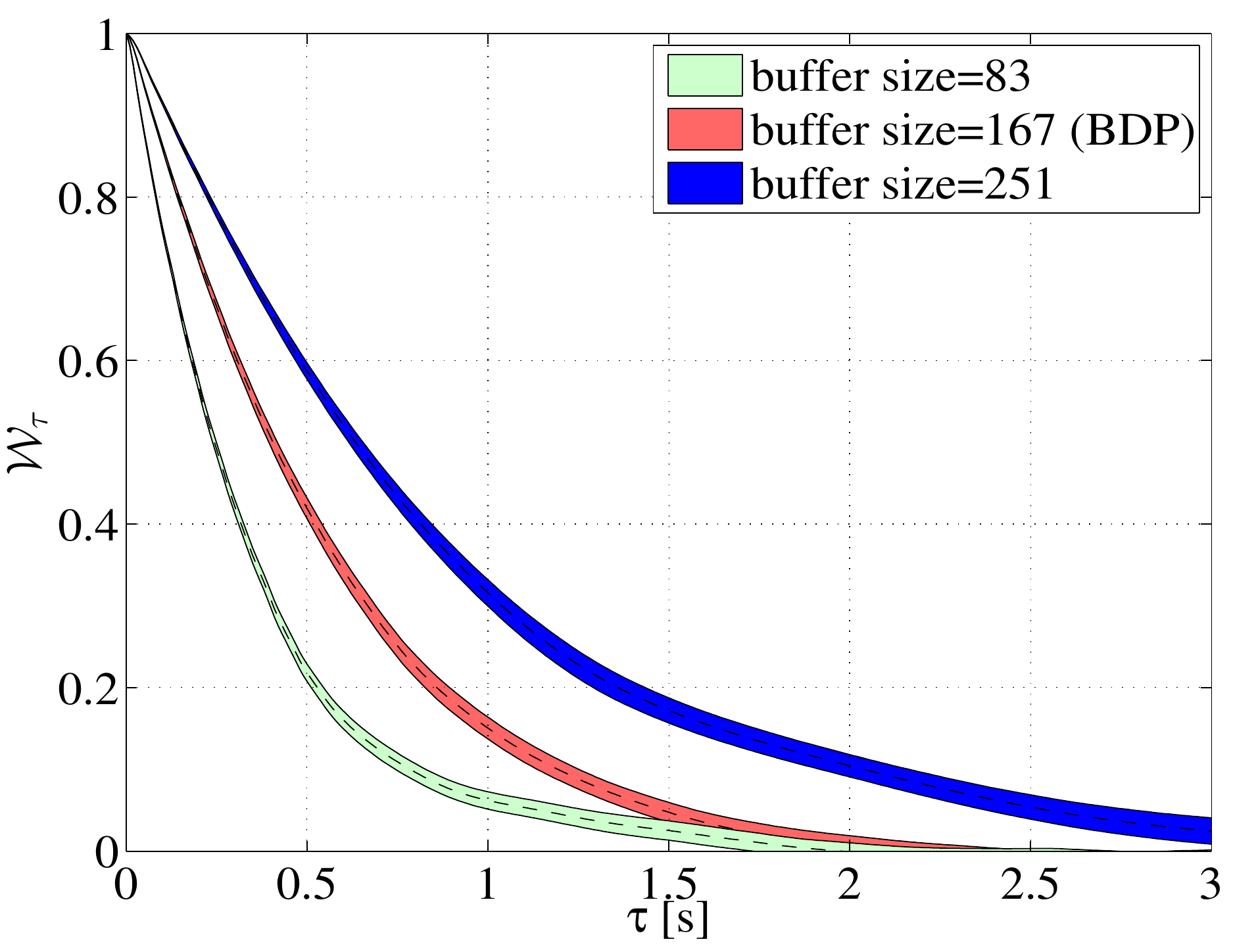}
    \label{fig:xcov}
  }
  \caption{Fig.~\protect\subref{fig:ccdf_queue_size} depicts CCDFs of the end-to-end delay for different router buffer sizes. The increase of the delays is explained by the autocorrelation of the CWND shown in Fig.~\protect\subref{fig:xcov}.}
  \label{fig:prob_stack_queueing_queue_size}
 \end{figure}

To understand how smaller router buffers favor the convergence of the attainable rate function, we consider the end-to-end delays for an arrival rate of 5~Mbit/s and buffer sizes of 83, 167, and 251 packets. The buffer sizes correspond to maximal queueing delays of 10~ms, 20~ms, and 30~ms, respectively. Fig.~\ref{fig:ccdf_queue_size} shows the CCDFs of the delays that, as expected, grow if the buffer size is increased. Compared to the network delays, the growth of the end-to-end delays is, however, larger by magnitudes, i.e., increasing the buffer size in routers causes a repercussion on the sender's TCP socket that exceeds the impact of the queueing delay notably. Interestingly, the growth applies to the tail of the distribution that is caused by queueing in the sender's protocol stack.

The implications of larger router buffers on stack delays are explained by a deceleration of the AIMD congestion window adaptation that is caused by increased RTTs. The effects of larger RTTs are twofold: firstly, a larger CWND is required to achieve the same throughput that is determined by CWND/RTT, and as a consequence the number of increments to recover the CWND after a multiplicative decrease becomes larger; and secondly, during additive increase the CWND is incremented by one MSS per RTT, so that the speed of the CWND growth is reduced. To quantify the effects, we use the autocorrelation of the CWND $\mathcal{W}_\tau$ at lag $\tau$ defined as~\cite{Grimmett2001}
\begin{equation}
\label{equ:acorr}
 \mathcal{W}_{\tau}= \frac{\mathsf{E}[(W(t)-\mu_W)(W(t+\tau)-\mu_W)]}{\sigma_W^2} ,
\end{equation}
where $\mu_W$ is the mean and $\sigma_W^2$ the variance of the CWND process $W(t)$. We note that $\mu_W$ and $\sigma_W^2$ increase with the buffer size in the network. Due to the normalization, the autocorrelation is, however, scale-free and enables a comparison.

The autocorrelation of the CWND for an arrival rate of 5~Mbit/s is presented in Fig.~\ref{fig:xcov}. In detail, the average of the autocorrelation obtained from 20 independent measurement runs is displayed. The colored areas indicate corresponding 0.95 confidence intervals. Generally, the results show a strong autocorrelation of the CWND over lags in the order of seconds, that is caused by the slow additive increase. The correlation increases notably with the buffer size.

The implication of strong correlations over large lags is that if a small CWND value occurs, it will likely prevail for long periods of time. The reduced transmission rate during these recovery times causes significant queueing in the sender's protocol stack that leads to large end-to-end delays. A remedy is to modify the adaptation rule of the CWND, i.e., the congestion control algorithm, as recent TCP versions do.

Concluding, the experiments show that oversized router buffers significantly reduce the performance on short and medium time-scales while undersized buffers can limit the long-term attainable rate. Given the unknowns in real networks, this makes the optimal configuration of router buffers difficult. AQM addresses this problem by adapting the packet loss probability.

\subsection{Impact of Active Queue Management}
\label{sec:aqm}
We compare the well-known random early detection (RED)~\cite{Floyd:1993} and the recently proposed controlled delay (CoDel)~\cite{nichols:codel12} AQM to drop-tail queueing with a BDP sized queue and TCP NewReno. RED drops packets randomly if the filling level of the buffer is between a minimal and a maximal threshold, whereby the drop probability increases linearly with the filling level. CoDel's drop strategy seeks to meet a target sojourn time of packets in the queue. In detail, we used adaptive RED introduced in~\cite{Floyd:adaptivered} with two configurations, the one described in~\cite{floyd:parameters}, and a second one, where the minimum threshold is configured to 20 packets instead of the default of 5 packets. For CoDel the default parameters were used. The AQM schemes were configured on the interfaces between the two routers. In addition, we compare the AQM schemes to synthetic loss processes that enable us to achieve a better understanding of the impact of the loss process. Using synthetic losses, we omit any cross traffic so packet loss only occurs due to the defined loss process.
\subsubsection{The Cost of Clustered Losses}
Fig.~\ref{fig:attain_rate_aqm} presents estimates of the attainable rate function $\widetilde{\alpha}^\varepsilon(t)$ for RED, CoDel, and drop-tail queueing, respectively. We used RED with a minimum threshold of 20 packets. All configurations achieve the target long-term rate of 9~Mbit/s, hence Fig.~\ref{fig:attain_rate_aqm} focuses on the short-term behavior only. Both RED and CoDel improve the short-term performance of TCP if compared to drop-tail queueing with a BDP sized queue. While RED achieves a better performance, it is sensitive to its parameters as already stated in~\cite{Floyd:adaptivered}. Conducting the experiment with the default minimum threshold of 5~packets~\cite{floyd:parameters}, limits the long-term rate to 8~Mbit/s instead of 9~Mbit/s. The default parameters of CoDel are suitable for this scenario.

Since the loss probability of RED and CoDel is dynamically adjusted to the utilization, the loss process is unknown. We also use well-defined synthetic loss processes for comparison  to infer the impact on the performance from the loss process. We use a loss process where loss occurs by independent Bernoulli trials and in periodic intervals, respectively. The loss probability is configured so that the same long-term throughput rate is achieved. Bernoulli losses and RED show a similar performance in the attainable rate, see Fig.~\ref{fig:attain_rate_synthetic_exp}. In contrast, periodic losses achieve a short-term rate which is nearly twice as large as in case of RED and in case of Bernoulli losses.
\subsubsection{Discussion of Loss Schemes}
\begin{figure}
\centering
  \subfloat[ $\widetilde{\alpha}^\varepsilon(t)$ - AQM ]{
   \includegraphics[width=0.47\columnwidth]{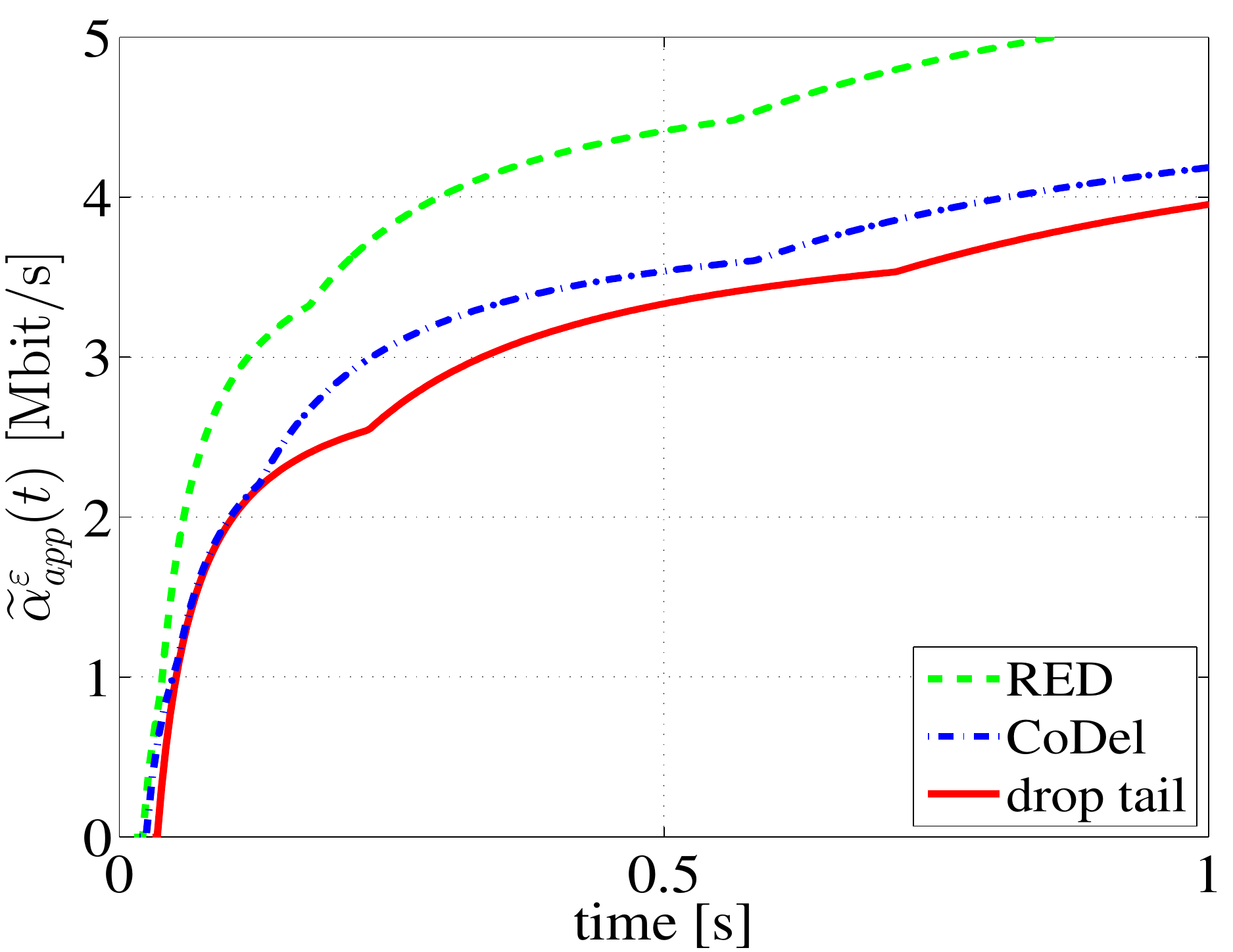}
   \label{fig:attain_rate_aqm}
 }
 \subfloat [ $\widetilde{\alpha}^\varepsilon(t)$ - synthetic loss ]{
   \includegraphics[width=0.47\columnwidth]{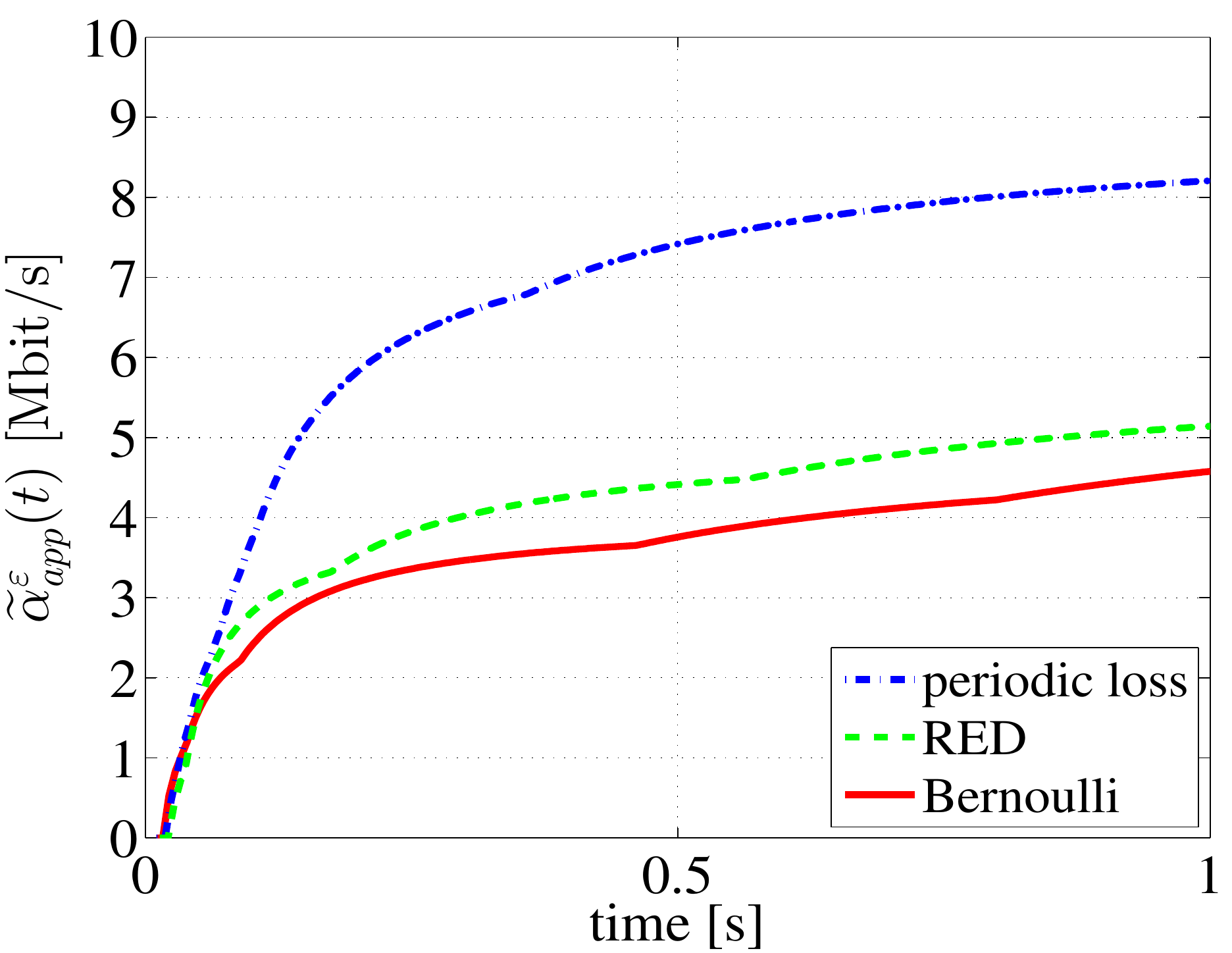}
   \label{fig:attain_rate_synthetic_exp}
 } \newline
  \subfloat[ autocorrelation - AQM]{
   \includegraphics[width=0.47\columnwidth]{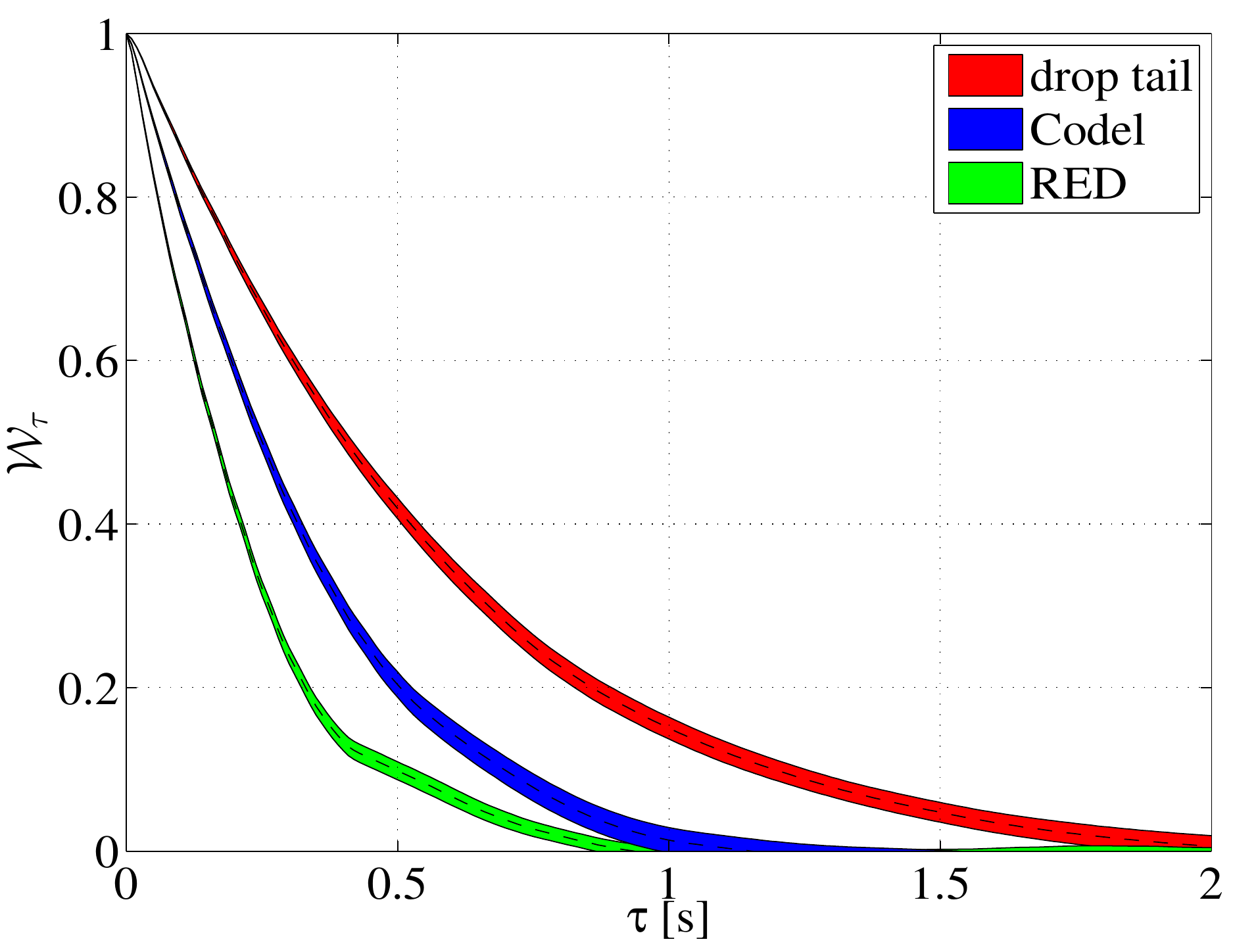}
   \label{fig:aqm_xcov}
 }
 \subfloat [ autocorrelation - synthetic loss ]{
   \includegraphics[width=0.47\columnwidth]{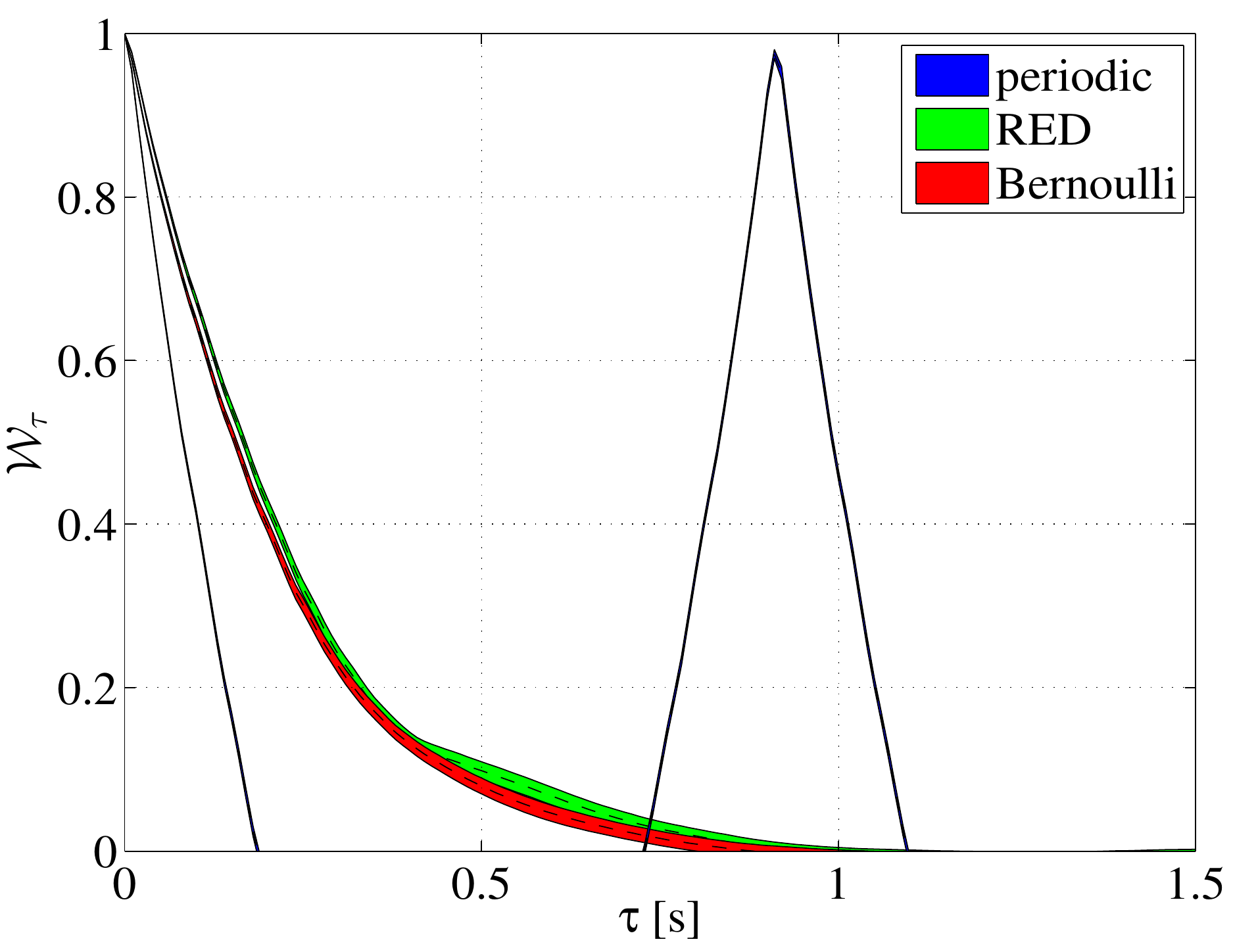}
   \label{fig:xcov_synthetic_exp}
 }
 \caption{Attainable rate and autocorrelation of the CWND for AQM schemes and synthetic loss processes. The autocorrelation is shown for a rate of 5~Mbit/s. The AQM schemes and synthetic loss processes, which avoid closely spaced losses, increase the attainable rate considerably.}
 \label{fig:aqm_random_loss}
\end{figure}
How the different AQM schemes and loss processes, respectively, improve TCP's performance can be explained by the characteristics of the CWND time series. In case of periodic losses, a significant performance improvement is observed. This is due to the fact that the CWND is completely recovered before the next loss occurs. As shown in Fig.~\ref{fig:xcov_synthetic_exp}, this leads to a faster decay of the CWND's autocorrelation than for Bernoulli losses and RED. Due to the periodicity of the losses, the autocorrelation shows a second peak at about 1~sec and repeatedly with this interval. Periodic losses achieve the best performance since the distance between packet losses is ideal. Any other loss process with the same loss probability would result in a higher correlation.

The effects are illustrated by the autocorrelation of the other loss schemes in Fig.~\ref{fig:aqm_xcov} and Fig.~\ref{fig:xcov_synthetic_exp}. RED and Bernoulli losses achieve a similar attainable rate and autocorrelation. Both schemes drop packets randomly and independently. Differences are in the loss process: in case of RED, the number of packets between two packet drops is uniformly distributed, to reduce the probability of closely spaced losses as described in~\cite{Floyd:1993}. Using CoDel the increase of the attainable rate is less than for RED, which is also reflected by the autocorrelation that spans across larger lags. CoDel determines the time to drop the next packet by a control law. In case of persistent queueing, the time interval to drop the next packet is decreased accordingly. Even though using a time interval to drop the next packets seems to be promising, since it would avoid closely spaced losses, in our scenario the attainable rate increases only marginally.

The difficulty of configuring the correct buffer size is avoided by AQM schemes that adapt to the utilization. Bernoulli losses and RED demonstrate that independent random losses improve the attainable rate significantly. However, improvements may be possible, as indicated by the periodic loss scheme.
\subsection{Impact of the Congestion Window Algorithm}
\label{sec:congestion_control}
The queue size and the queue management decide when to drop a packet. The reaction to a packet drop is determined by the congestion control algorithm. The main algorithms differ from each other primarily by the increase of the CWND after acknowledgements of successfully transmitted packets and the decrease after the detection of a packet loss if in congestion avoidance. Well-known congestion control protocols are TCP NewReno, TCP Cubic, and TCP Scalable. TCP NewReno was the default algorithm for many years. TCP Cubic and TCP Scalable are replacements that aim at achieving better performance in networks with large BDPs. Generally, oscillations of the window size cause variations of the transmission rate. The congestion control algorithm and the network path establish a closed control loop with mutual influences. To evaluate the performance of TCP NewReno, TCP Cubic, and TCP Scalable~\cite{Kelly:Scalable} we estimate their attainable rate functions $\widetilde{\alpha}^\varepsilon(t)$.
\begin{figure}
 \centering
  \subfloat[ $\widetilde{\alpha}^\varepsilon(t)$ ]{
    \includegraphics[width=0.47\columnwidth]{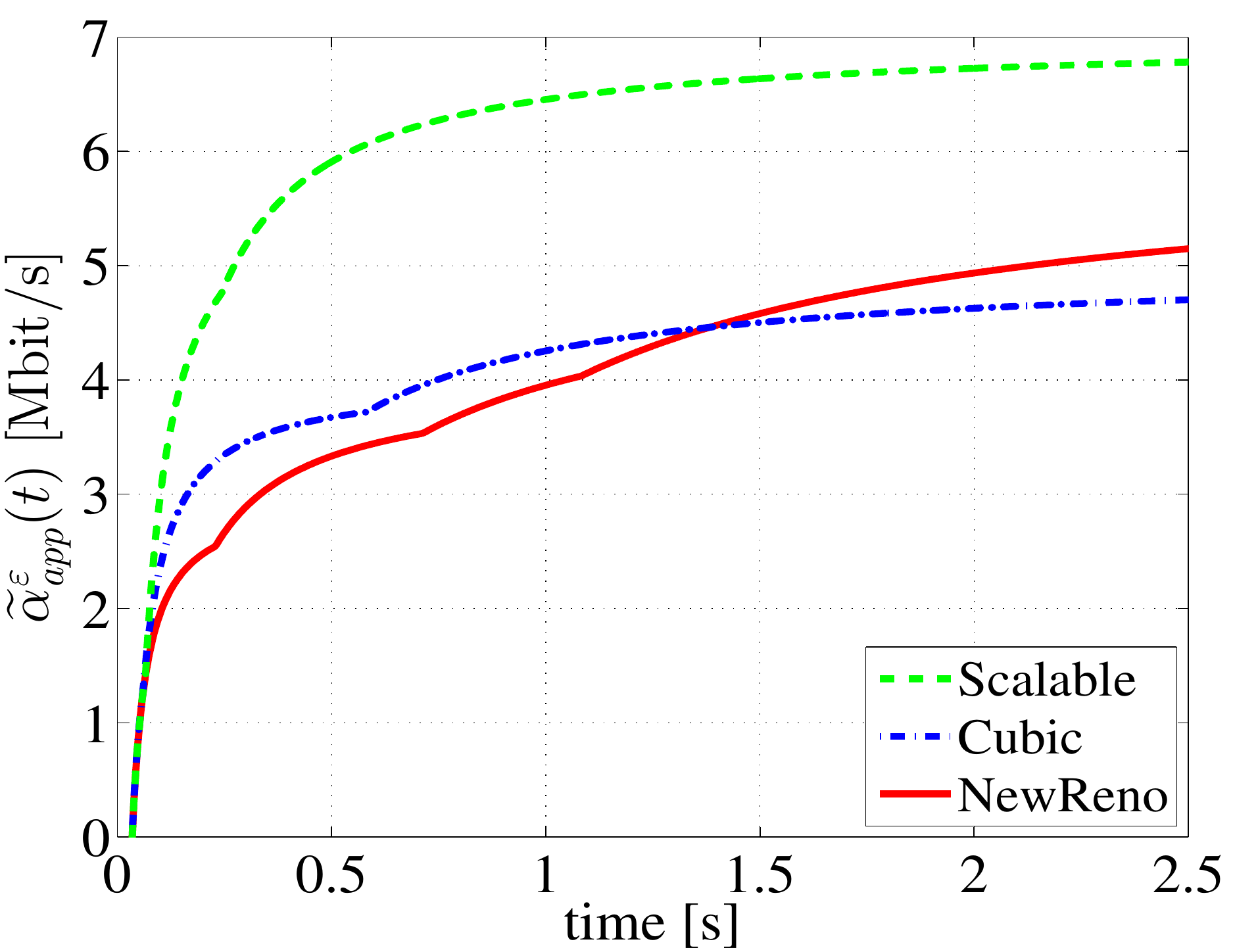}
    \label{fiq:cong_st}
  }
   \subfloat [ autocorrelation of the CWND ]{
     \includegraphics[width=0.47\columnwidth]{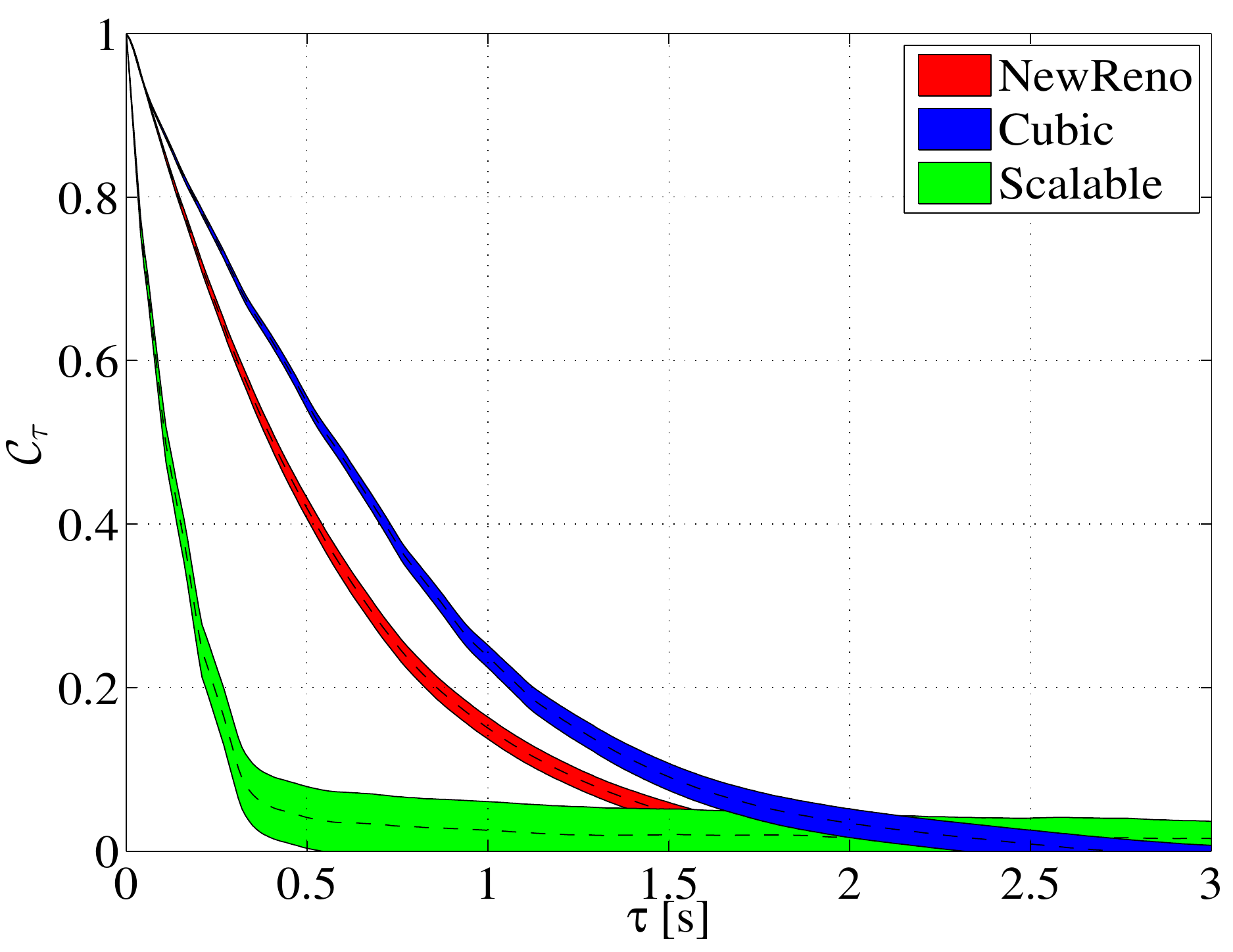}
     \label{fiq:cong_xcov}
   }
  \caption{Estimates of the attainable rate function and the autocorrelation for different congestion control algorithms. The algorithm Scalable that implements a multiplicative-increase multiplicative-decrease rule \cite{Mills10studyof}, increases the performance.}
  \label{fig:aqm_cc_st}
\end{figure}

We use the topology depicted in Fig.~\ref{fig:network_tcp} and drop-tail queueing with a queue size equal to the BDP at the routers. In each experiment, all nodes use the same congestion control protocol. All of the three protocols achieve a long-term rate of $9$~Mbit/s. The estimates $\widetilde{\alpha}^\varepsilon(t)$ exhibit, however, significant differences on short time-scales that are depicted in Fig.~\ref{fiq:cong_st}. TCP NewReno and TCP Cubic achieve a similar rate, where TCP Cubic has a slightly better performance for short time intervals of up to one second. Both algorithms are outperformed by TCP Scalable. This is a result of the different algorithms for adaptation of the CWND. In case of a loss, TCP Scalable reduces the CWND to 7/8 of the previous CWND value and subsequently it increases the CWND by one for each 100 packets. This leads to a faster recovery as illustrated in~\cite{Kelly:Scalable}. TCP NewReno and TCP Cubic reduce the CWND more than Scalable namely to 1/2 CWND, respectively, 4/5 CWND. TCP  NewReno increases the CWND linearly by one during each RTT. TCP Cubic uses a cubical function, which increases fast immediately after a loss, yet the increase slows down gradually. This adaptation achieves fairness with respect to TCP NewReno, which is not ensured by TCP Scalable as reported in~\cite{Li:evaluation}. Nevertheless, the quick recovery from a packet loss that is achieved by TCP Scalable significantly reduces the autocorrelation of the CWND depicted in Fig.~\ref{fiq:cong_xcov} and improves the performance accordingly.
\section{Comparison of Application Level Delays}
\label{sec:comparison}
An essential performance measure for real-time applications is the end-to-end delay. Using the characterization of the end-to-end connection by the attainable rate, respectively, the service curve, delay bounds follow for given arrival traffic characteristics. Here, we present the end-to-end delay bound for a video traffic trace to evaluate the various TCP options that we estimated before. We determine the end-to-end delay from the service curve estimate from Eq.~\eqref{equ:sc_estimate} and an empirical video envelope, see~\cite{wrege:vbr}, that is generated from a data trace of an H.264 encoded movie with a mean rate of 4~Mbit/s. This rate represents a moderate utilization with respect to the long-term attainable rate of about 9.4 Mbit/s.
\begin{figure}
    \centering
    \includegraphics[width=0.8\columnwidth]{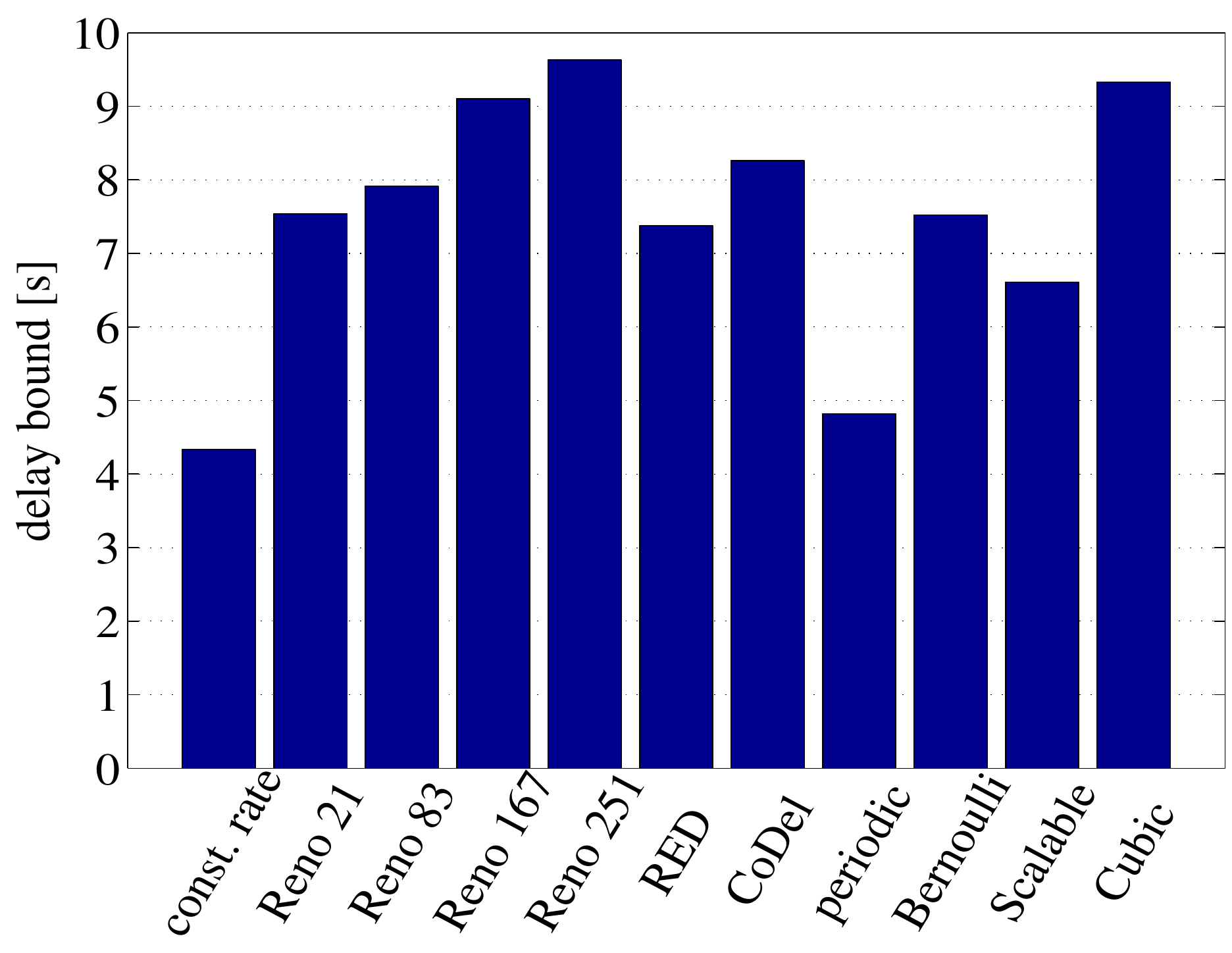}
    \caption{End-to-end delay bounds computed from service curve estimates for the TCP configurations determined in the previous sections and the empirical envelope for a video data trace with a mean rate of 4~Mbit/s.}
    \label{fig:trace_delay}
\end{figure}

Fig.~\ref{fig:trace_delay} shows the end-to-end delays for $\xi=0.01$. First, we depict the delay for a transmission of the video by a system that offers a constant service rate of 9.4~Mbit/s without any losses. We use this result as a baseline for comparison, as delays are only due to the variability of the video traffic in this case. Interestingly, the delay bound that applies in case of periodic losses is close to the delay bound for the constant rate service. If packet losses occur randomly, the delay bound increases. Drop-tail queues with small queue size and random loss schemes as RED and losses due to a Bernoulli process achieve a similar bound. TCP NewReno was used for all these before mentioned experiments. Also, changing the congestion control algorithm can achieve a much better delay bound as illustrated for TCP Scalable, which recovers the window faster after a loss. TCP Cubic and TCP NewReno perform comparably in our experiments.
\section{Conclusions}
\label{sec:conclusion}
\balance
We contributed an estimation method for systematic evaluation of the delay performance of closed-loop flow control protocols. The method is based on a model that characterizes flow control by a random service process. Estimates of this service process are obtained from end-to-end delay measurements. Applied to the prevalent closed-loop flow control protocol TCP, the estimation method reveals that the attainable rate on short to medium time-scales is significantly below the long-term rate. A decomposition of end-to-end delays showed that closed-loop congestion control accounts for the largest part of the delays. We found that the CCDF of end-to-end delays is composed of two distinctive regions, that are due to the network and the sender's protocol stack, respectively. Noteworthy, large delays arise in the sender's protocol stack already for data rates that are much below the long-term attainable rate. Using the estimation method for a comparative evaluation, we demonstrate the considerable impact of certain parameters important for network and server dimensioning on the delay performance. In detail, we observe a strong connection between utilization and delay. Furthermore, we discovered that the amount of queueing in the sender's protocol stack increases with the buffer size that is configured in routers. We showed that this effect is due to the correlation of the CWND, that grows with the buffer size. Moreover, our evaluation showed that AQM and TCP Scalable can lead to less correlation of the CWND process and thus improve the delay performance. The evaluation concluded with a comparison of the application level delay for transmission of a video stream to evaluate the performance for actual traffic arrivals.
\bibliographystyle{IEEEtran}
\bibliography{../bibliography/IEEEabrv,../bibliography/rfc,../bibliography/tcp,../bibliography/bandwidthestimation,../bibliography/networkcalculus,../bibliography/misc}
\end{document}